\documentclass[apj]{emulateapj}
\usepackage{comment} 
\usepackage{enumerate}
\usepackage{lscape}

\shorttitle{STAR FORMATION AND AGN FRACTIONS AT HIGH REDSHIFT}
\shortauthors{MURPHY ET AL.}
\slugcomment{Draft version 1.5  April 6, 2009}
\journalinfo{Accepted to \apj}

\begin{document}

\title{Balancing the Energy Budget between Star-Formation and AGN in High Redshift Infrared Luminous Galaxies}

\author{E.J.~Murphy,\altaffilmark{1} R.-R. Chary,\altaffilmark{1} 
D.M.~Alexander,\altaffilmark{2} 
M.~Dickinson,\altaffilmark{3} 
B.~Magnelli,\altaffilmark{4} 
G.~Morrison,\altaffilmark{5,6} 
A.~Pope\altaffilmark{3,7}, H. I. Teplitz\altaffilmark{1}}

\altaffiltext{1}{\scriptsize {\it Spitzer} Science Center, MC 314-6, California Institute of Technology, Pasadena, CA 91125; emurphy@ipac.caltech.edu}
\altaffiltext{2}{\scriptsize Department of Physics, Durham University, Durham DH1 3LE, UK}
\altaffiltext{3}{\scriptsize National Optical Astronomy Observatory, Tucson, AZ 85719}
\altaffiltext{4}{\scriptsize Laboratoire AIM, CEA/DSM-CNRS-Universit\'{e} Paris Diderot, DAPNIA/Service d'Astrophysique, B\^{a}t. 709, CEA-Saclay, F-91191 Gif-sur-Yvette C\'{e}dex, France}
\altaffiltext{5}{\scriptsize  Institute for Astronomy, University of Hawaii, Honolulu, HI 96822}
\altaffiltext{6}{\scriptsize Canada-France-Hawaii Telescope, Kamuela, HI 96743}
\altaffiltext{7}{\scriptsize {\it Spitzer} Fellow}

\begin{abstract}
We present deep {\it Spitzer} mid-infrared spectroscopy, along with 16, 24, 70, and 850\,$\micron$\ photometry, for 22 galaxies located in the Great Observatories Origins Deep Survey-North (GOODS-N) field.  
The sample spans a redshift range of $0.6\la z \la 2.6$, 24~$\mu$m flux densities between $\sim$0.2$-$1.2 mJy, and consists of submillimeter galaxies (SMGs), X-ray or optically selected active galactic nuclei (AGN), and optically faint ($z_{AB}>25$\,mag) sources. 
We find that infrared (IR; $8-1000~\micron$) luminosities derived by fitting local spectral energy distributions (SEDs) with 24~$\micron$ photometry alone are well matched to those when additional mid-infrared spectroscopic and longer wavelength photometric data is used for galaxies having $z\la1.4$ and 24~$\micron$-derived IR luminosities typically $\la 3\times 10^{12}~L_{\sun}$.  
However, for galaxies in the redshift range between $1.4\la z \la 2.6$, typically having 24~$\micron$-derived IR luminosities $\ga 3\times 10^{12}~L_{\sun}$, IR luminosities are overestimated by an average factor of $\sim$5 when SED fitting with 24~$\micron$ photometry alone.  
This result arises partly due to the fact that high redshift galaxies exhibit aromatic feature equivalent widths that are large compared to local galaxies of similar luminosities.  
Using improved estimates for the IR luminosities of these sources, we investigate whether their infrared emission is found to be in excess relative to that expected based on extinction corrected UV star formation rates (SFRs), possibly suggesting the presence of an obscured AGN.  
Through a spectral decomposition of mid-infrared spectroscopic data, we are able to isolate the fraction of IR luminosity arising from an AGN as opposed to star formation activity.   
This fraction is only able to account for $\sim$30\% of the total IR luminosity among the entire sample and $\sim$35\% of the ``excess" IR emission among these sources, on average, suggesting that AGN are not the dominant cause of the inferred ``mid-infrared excesses" in these systems.  
Of the sources identified as having mid-infrared excesses, half are accounted for by using proper bolometric corrections while half show the presence of obscured AGN.  
This implies sky and space densities for Compton-thick AGN of $\sim$1600~deg$^{-2}$ and $\sim$$1.3\times 10^{-4}$~Mpc$^{-3}$, respectively.   
We also note that IR luminosities derived from SED fitting the mid-infrared {\it and} 70~$\micron$ broadband
photometry agree within $\sim$50\% to those values estimated using the additional mid-infrared spectroscopic and submillimeter data.  
An inspection of the FIR-radio correlation shows no evidence for evolution over this redshift range.   
However, we find that the SMGs have IR/radio ratios which are a factor of $\sim$3 lower, on average, than what is measured for star-forming galaxies in the local Universe.  

\end{abstract}

\section{Introduction}

A reliable estimation of the star formation rates (SFRs), stellar mass, and active galactic nuclei (AGN) activity at all redshifts is required in order to obtain a comprehensive understanding of galaxy evolution.  
Various studies have shown that the SFR density increases by a factor of $\sim$5$-$10 between $z\sim0$ and $z\sim1$ and that $\gtrsim$60\% of the total SFR density at $z\sim1$ is obscured by dust \citep{de02,ce01,el05,bm09}.  
Until now, these results have relied on deep  surveys at 15\,$\micron$ and 24\,$\micron$ which sample the redshifted mid-infrared spectral energy distribution (SED) of galaxies.

The mid-infrared ($5-40~\micron$) SED is a complex interplay of broad emission features thought to arise from polycyclic aromatic hydrocarbon (PAH) molecules, silicate absorption features at 9.7 and 18~$\micron$, and a mid-infrared continuum from very small grains \citep{lp84,atb85}.
Mid-infrared luminosities measured near $\sim$8~$\micron$ have been found to correlate with the total infrared (IR; $8-1000$) luminosities of galaxies in the local Universe \citep{ce01, de02}, which itself is a measure of a galaxy's SFR.  
While a correlation is found, there does exist a large amount of scatter \citep{dd05,jd07,la07} and systematic departures for low metallicity systems \citep{ce05,sm06}.  
It is therefore uncertain if these empirical relations between mid-infrared and total IR luminosities ($L_{\rm IR}$) for galaxies in the local Universe are applicable for galaxies at higher redshifts.  

Recently, data from deep far-infrared surveys such as the Far-infrared Deep Legacy Survey (FIDEL; PI: M. Dickinson) are confirming the same redshift evolution of the SFR density seen in the mid-infrared and do not show any evolution in the SED of infrared luminous galaxies \citep[e.g.][]{bm09}.
Furthermore, individual lensed Lyman break galaxies (LBGs) at $z\sim2.5$ \citep{bs08, gw08, ag08}, seem to show bolometric corrections similar to those of local galaxies.  
In contrast, \citet{jr08} have provided evidence that the high redshift lensed mid-infrared selected galaxies might show a factor of $\sim$2 stronger rest-frame 8~$\micron$ emission compared to their total IR luminosities.
Part of the contradiction might be due to the intrinsic scatter in the physical properties of the galaxies, particularly
the range of dust color temperatures and emissivities seen for galaxies in the local Universe \citep[e.g.][]{ld00}.  
Individual galaxies detected in the far-infrared or submillimeter at high redshift, for which the comparisons are made, might be sampling the scatter in the ratio rather than the median trend which the empirical galaxy templates follow.

In order to assess the accuracy of mid-infrared derived bolometric luminosities as a tracer of star-formation at $z\sim2$,
\citet{ed07a,ed07b} compared mid-infrared (i.e. observed-frame 24~$\micron$) based SFRs to those derived from extinction corrected UV emission and radio emission.
They found a systematic offset, whereby the mid-infrared derived bolometric luminosities were systematically higher than alternate SFR estimates, especially among the most intrinsically luminous mid-infrared sources. 
Furthermore, they found that stacking the X-ray data on these ``mid-infrared excess" galaxies revealed the presence of a hard X-ray source suggestive of obscured AGN activity. Since mid-infrared wavelengths are sensitive to warm dust continuum emission arising from AGN activity, \citet{ed07b} concluded that these mid-infrared excesses arise from deeply obscured AGN and report a space density of Compton thick AGN which is twice that of all X-ray detected AGN at $z\sim2$.

The detection of a source in hard X-rays does not necessarily imply that an AGN is dominating its energetics as is found for submillimeter galaxies (SMGs) which appear to be both star-formation driven and commonly detected in the X-rays \citep{da05, ap08a}.
Furthermore, evidence for ``mid-infrared excess" relative to extinction corrected UV luminosities might simply be evidence for optically thick star-formation, a phenomenon which is ubiquitous among ultraluminous infrared galaxies (ULIRGs; $L_{\rm IR} \geq 10^{12}~L_{\sun}$) in the local Universe \citep{bv07}.  
Mid-infrared spectroscopy of galaxy samples selected by their mid-infrared flux densities provides a powerful technique to isolate these various selection effects among infrared luminous sources.
By decomposing these spectra, into PAH and continuum components, one can get a handle on the relative fraction of power being emitted by AGN activity within such sources relative to the output related to star formation \citep[e.g.][]{as07,ap08a}.  

The Great Origins Observatories Deep Survey-North \citep[GOODS-N;][]{md03} field, due to the wealth of multiwavelength data spanning the ultraviolet to radio wavelengths, is the optimal field in which to undertake such a relatively unbiased survey.
Using mid-infrared spectroscopy for 22 sources selected at 24~$\micron$ in the GOODS-N field, along with existing 850~$\micron$ and additional 70~$\micron$ imagery obtained as part of FIDEL, 
we aim to improve our understanding of the star formation and AGN activity within a diverse group of $0.6\la z \la2.6$ galaxies.  
This is done through a more proper estimate of IR luminosities and the ability to decompose these measurements into star-forming and AGN components.  

The paper is organized as follows: 
In $\S$2 we introduce the sample and describe the observations.  
The determination of IR luminosities from SED fitting, calculation of SFRs, and the decomposition of AGN and star-formation power using the mid-infrared spectroscopy are described in $\S$3.  
The results are presented in $\S$4 while their implications are discussed in $\S$5.  
Finally, in $\S$6, we summarize our conclusions.  

\section{The Spectroscopic Sample and Multiwavelength Photometry}
The mid-infrared spectroscopic sample presented here
consists of 22 sources drawn from 24~$\micron$ detections in the Great Observatories Origins Deep Survey-North \citep[GOODS-N;][]{md03} field.  
Galaxy coordinates and IDs used throughout the text are given in Table~\ref{tbl-1}.  
The sample was chosen to span a range in redshift from $0.6 \la z \la 2.6$ and contain a variety of object types: SMGs, X-ray or optically selected AGN, typical luminous infrared galaxies (LIRGs; $10^{11} \leq L_{\rm IR} < 10^{12}~L_{\sun}$),  and optically faint (i.e. $z_{\rm AB} > 25$ mag) 24~$\micron$ sources. 
The classification for each object is given in Table~\ref{tbl-2}.  
To ensure high signal-to-noise in the IRS spectra, a flux density cut of $f_{\nu}(24~\micron) > 200~\mu$Jy was made,  although the median flux density
of the sample is 510~$\mu$Jy.  
The multiwavelength photometry is given in Table \ref{tbl-3}.  

\begin{deluxetable}{cccccc}
\tablecaption{Galaxy Positions and Redshifts \label{tbl-1}}
\tablewidth{0pt}
\tablehead{
\colhead{} & \colhead{R.A.} & \colhead{Decl.} & \colhead{} & \colhead{} & \colhead{$z_{\rm opt}$}\\
\colhead{ID} & \colhead{(J2000)} & \colhead{(J2000)} & \colhead{$z_{\rm opt}^{a}$} & \colhead{$z_{\rm IRS}^{b}$} & \colhead{Reference}
}
\startdata
  1&  12~35~55.13&  62~ 9~ 1.7&    1.875&  1.88&   1\\
  2&  12~36~ 0.17&  62~10~47.3&  2.002&  2.01&   1\\
  3&  12~36~ 3.25&  62~11~10.8&  0.638&  0.63&   2\\
  4&  12~36~16.11&  62~15~13.5&  2.578&  2.55&   1\\
  5&  12~36~18.33&  62~15~50.4&  1.865&  2.00&   1\\
  6&  12~36~19.13&  62~10~ 4.3&    \nodata&  2.21&   \nodata\\
  7&  12~36~21.27&  62~17~ 8.1&  1.992&  1.99&   3\\
  8&  12~36~22.48&  62~15~44.3&  0.639&  0.64&   2\\
  9&  12~36~22.66&  62~16~29.5&  2.466&  1.79&   1,3\\
 10&  12~36~33.22&  62~ 8~34.7&  0.934&  0.94&   2\\
 11&  12~36~34.51&  62~12~40.9&  1.219&  1.22&   4\\
 12&  12~36~37.02&  62~ 8~52.2&    \nodata&  2.01&   \nodata\\
 13&  12~36~53.22&  62~11~16.7&  0.935&  0.93&   5\\
 14&  12~36~55.93&  62~ 8~ 8.1&  0.792&  0.79&   5\\
 15&  12~36~56.47&  62~19~37.5&  1.362&  2.20&   6\\
 16&  12~37~ 1.59&  62~11~46.2&  0.884&  1.73&   7\\
 17&  12~37~ 4.34&  62~14~46.1&  2.211&  2.21&   8\\
 18&  12~37~11.37&  62~13~31.1&  1.996&  1.98&   3\\
 19&  12~37~16.59&  62~16~43.2&    \nodata&  1.82&   \nodata\\
 20&  12~37~18.27&  62~22~58.9&    \nodata&  1.88&   \nodata\\
 21&  12~37~26.49&  62~20~26.6&    \nodata&  1.75&   \nodata\\
 22&  12~37~34.52&  62~17~23.2&  0.641&  0.64&   5
      \enddata
    \tablecomments{$^{a}$Optical spectroscopic redshift. $^{b}$Mid-infrared (IRS) spectroscopic redshift.} 
    \tablerefs{(1)~\citep{sc05}; (2)~\citet{gw04}; (3)~\citet{as04}; (4)~\citet{jc96}; (5)~\citet{lc04}; (6)~D. Stern et al. 2009, in preparation; (7)~\citet{jc00}; (8)~\citet{cs02}}
\end{deluxetable}

\subsection{IRS Spectroscopy}
\begin{figure*}
\plotone{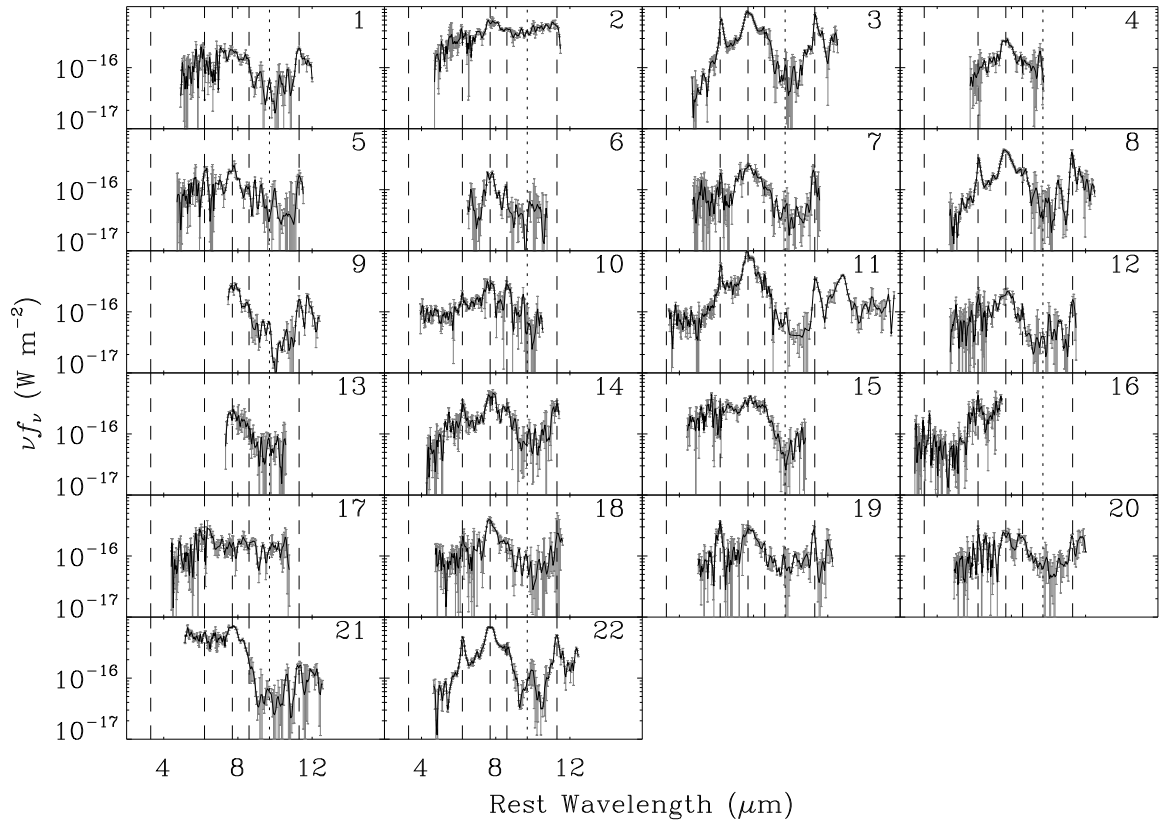}
\caption{The mid-infrared IRS spectra for each of the 22 sample galaxies.  
Each spectra is shown with associated errors in the rest-frame and has been smoothed to the instrument resolution.  
The expected positions of the 3.3, 6.2, 7.7, 8.6, and 11.3~$\micron$ PAH features are indicated by {\it dashed} vertical lines; the expected position of the 9.7~$\micron$ silicate absorption feature  is indicated with a {\it dotted} vertical line.  
\label{fig-1}}
\end{figure*}

{\it Spitzer} IRS observations were taken as part of {\it Spitzer} program GO-20456 (PI: R.-R. Chary) in April/May 2006.    
These low resolution ($R = \lambda/\Delta \lambda \sim 100$) spectroscopic observations were made using the spectral staring mode, which observed each target at two nod positions (per cycle) within the slit.  
In Table~\ref{tbl-2} we provide the ramp times and the total number of cycles used per low-resolution module.  
We note that a study focusing on the SMGs included in this sample can be found in \citet{ap08a} where complete details on the observation strategy and data reduction steps are also presented.   
The actual spectra, shifted into the rest frame, are shown in Figure \ref{fig-1} along with associated errors.  
The expected positions of the 3.3, 6.2, 7.7, 8.6 and 11.3~$\micron$ PAH features, along with the the 9.7~$\micron$ silicate absorption feature, are indicated with vertical lines.    

For cases where PAH features in the IRS spectra are clearly identified, redshifts were measured 
directly as described in \citet{ap08a}.  
The redshifts for the remaining galaxies were taken from existing optical spectroscopy (Table \ref{tbl-1}).  
We note that there were some discrepancies in the IRS and optical spectroscopic redshifts (i.e. GN-IRS~5,9,15, and 16) in which case the IRS-derived redshifts were used. 
GN-IRS~5 and 9 are both SMGs for which the discrepancies between their IRS and optical redshifts are discussed in Appendix~B of \citet{ap08a}.  
\citet{ap06} report that the optical ($z=0.884$) redshift associated with GN-IRS~16, also an SMG, is not the true counterpart.   
The discrepancy for GN-IRS~15 is thought to arise from a superposition of sources.  
The optical morphology of this source is quite complex, having several components and a range of colors.  
Thus, we strongly suspect we are seeing a superposition of a mid-infrared--bright galaxy located at $z=2.20$ and a foreground galaxy at $z=1.362$.  

The final redshift range among the sample spans $0.6 \la z \la 2.6$ with a median of 1.88.  
A total of 15 galaxies have a redshift between $1.4 \la z \la2.6$, which is almost exactly the nominal  redshift range for the $BzK$ color selection where a comparison with the \citet{ed07b} results are appropriate.  
In addition, 7 of the galaxies have redshifts between $0.6\la z \la 1.2$ and are intended to be representative of the LIRG population.  

\subsection{Mid- and Far-Infrared {\it Spitzer} Imaging}
Infrared imaging of the GOODS-N field was obtained at 16, 24, and 70~$\micron$ using the $Spitzer$ Space Telescope.  
The 24~$\micron$ observations of GOODS-N were taken as part of the GOODS legacy program (PI: Dickinson) and reach an RMS of $\sim$5~$\mu$Jy \citep[see][]{rc07}.   
The spectroscopic sample presented here spans the range $220-1220$\,$\mu$Jy. 

IRS peak-up observations at 16~$\micron$ \citep{ht05} were made covering an area slightly less extended than the GOODS-N 24~$\micron$ map.  
Only one of the sample galaxies, GN-IRS~1, lies outside the areal coverage of the 16~$\micron$ map.  
These 16~$\micron$ data  have an RMS of $\sim$6~$\mu$Jy and only sources with a signal-to-noise (S/N) ratio greater than 5 were considered as a 
detection.
Of the 21 sources in the 16~$\micron$ image field-of-view, 2 were not detected and their flux densities were set to the upper limit value of 30~$\mu$Jy.   
The 16~$\micron$ flux densities among the detected sources span a factor of $\sim$25, ranging from $\sim$$40-1050~\mu$Jy with a median value of $\sim$280~$\mu$Jy.  

MIPS 70~$\micron$ observations were carried out as part of program GO-3325 \citep[PI:][]{df06} and the Far Infrared Deep Extragalactic Legacy project (FIDEL; PI: Dickinson).  
The full field extends beyond the 24~$\micron$ coverage and data processing details can be found in \citet{df06}.  
The typical point-source noise of the 70~$\micron$ data is $\sim$0.55~mJy \citep{df06}, $\sim$1.6 times larger than the confusion noise level of $\sigma_{\rm c} = 0.35\pm0.15$~mJy estimated by \citet{df09}.  
Sources having a flux density $>~2~$mJy and a S/N$~>6$ were considered detections, leading to a total of 9 sources; all other sources were assigned an upper limit of 3~mJy since this value corresponds to a differential completeness of $\sim$80\% \citep{bm09}.     
The median 70~$\micron$ flux density among the detected sources is $\sim$5.5~mJy and ranges between $\sim$$2.2-14.2~$mJy, spanning a factor of roughly 6.  
The photometric uncertainty including calibration uncertainties and confusion is $\sim$10\% at 16 and 24~$\micron$ and $\sim$20\% at 70~$\micron$.     

\subsection{X-ray Imaging}
X-ray imaging of GOODS-N was obtained with the {\it Chandra} X-ray Observatory for a 2~Ms exposure \citep{da03}.  
The full band ($0.5-8.0$~keV) and hard band ($2.0-8.0$~keV) on-axis sensitivities are $\sim$$7.1\times10^{-17}$ and $1.4\times10^{-16} {\rm erg~cm^{-2}~s^{-1}}$, respectively.   
Of the 22 sample galaxies, 12 were matched with an X-ray counterpart reported by \citet{da03}.  
For the remaining 10 sources, upper limits were measured assuming $\Gamma=1.4$.

Among the 12 X-ray detected sources we find that the $0.5-8.0$~keV flux spans more than 2 orders of magnitude, ranging from $9-1120\times10^{-17}~{\rm erg~cm^{-2}~s^{-1}}$, with a median flux of $56\times10^{-17}~{\rm erg~cm^{-2}~s^{-1}}$ (Table \ref{tbl-3}).  
Eight sources have secure detections in the hard band ($2.0-8.0$~keV) while only upper limits could be measured for the remaining 14 sources.
Of those sources detected, the range in the $2.0-8.0$~keV flux spans between $25-1050\times10^{-17}~{\rm erg~cm^{-2}~s^{-1}}$ (i.e. a factor of $\sim$40) with a median flux of $\sim$$100\times10^{-17}~{\rm erg~cm^{-2}~s^{-1}}$.  

\subsection{Optical Imaging}
The GOODS-N field has been imaged extensively in four bands, $B_{435},~V_{606},~ i_{775}, ~{\rm and}~z_{850}$ using the ACS camera on the {\it Hubble} Space Telescope \citep{mg04}.  
Additional ground-based imaging of GOODS-N in the $U$-band was obtained using the prime-focus MOSAIC camera on the KPNO Mayall 4~m telescope \citep{mg04, cap04}.  
The 3-$\sigma$ upper limits of the $UBViz$ photometry are 0.027, 0.048, 0.048, 0.091, and 0.131~$\mu$Jy, respectively.

\subsection{Radio Continuum Imaging}
The GOODS-N field was observed with the Very Large Array (VLA) for a total of 165 hours using all 4 configurations. (G. Morrison et al. 2008, in preparation). 
Given that the VLA is a centrally condensed array, the observing time per configuration was scaled as follows, A-array (1), B-array (1/4), C-array (1/16), and D-array (1/64) \citep{fo08}.  
This integration time scaling provides the best sensitivity for larger sources.  
The final radio map has a local RMS at the phase center of $\sim$4~$\mu$Jy.  
All sources in the IRS spectroscopic sample were matched with a S/N$>4$ radio source to a radius of 1$\arcsec$ except for 
GN-IRS~13,  which had a S/N$\sim$3.3.  
The 20~cm flux densities span a factor of $\sim$9 (i.e. $\sim$$21-202~\mu$Jy) with a median value of $\sim$98~$\mu$Jy.  

\subsection{Submillimeter Imaging}
The GOODS-N field was imaged at 850~$\micron$ by a number of programs using the Submillimeter Common-User Bolometer Array (SCUBA) and the data were combined into the SCUBA ``supermap'' \citep{cb03,ap05}.  
The 8 SMGs targeted for IRS spectroscopy were selected from the ``supermap'' sources with robust identifications \citep{ap06} in addition to 3 SCUBA photometry sources from \citet{sc05}. 
Only 2 of the 11 SMGs are detected at 70~$\micron$ (Huynh et al. 2007). 
The 850~$\micron$ flux densties span a range between  $2-9$~mJy with a median flux density $5.2$\,mJy.
For the remaining non-detected IRS targets, 3-$\sigma$ upper limits were calculated from the 850~$\micron$ SCUBA supermap. 
The only exception is GN-IRS~20 which lies outside the areal coverage of the SCUBA maps.  

\begin{deluxetable}{ccccc}
\tablecaption{{\it Spitzer} IRS Observations Summary \label{tbl-2}}
\tablewidth{0pt}
\tablehead{
\colhead{} &
\multicolumn{3}{c}{Number of Cycles per Module$^{a}$} &
\colhead{}\\
\cline{2-4}\\
\colhead{} & 
\colhead{SL1} & \colhead{LL1} & \colhead{LL2} & \colhead{Selection}\\
\colhead{ID} & 
\colhead{(240~s)$^{b}$} & \colhead{(120~s)$^{b}$} & \colhead{(120~s )$^{b}$} &
\colhead{Criteria$^{c}$}
}
\startdata
  1&  \nodata&   40&   31&     OF,X-ray,SMG\\
  2&  \nodata&   10&    6&             SMG\\
  3&   10&  \nodata&    6&      X-ray,LIRG\\
  4&  \nodata&   70&  \nodata&     OF,X-ray,SMG\\
  5&  \nodata&   55&   35&           OF,SMG\\
  6&  \nodata&   55&  \nodata&           OF,SMG\\
  7&  \nodata&   45&   35&           OF,SMG\\
  8&   15&  \nodata&   12&      X-ray,LIRG\\
  9&  \nodata&   45&  \nodata&       X-ray,SMG\\
 10&   10&  \nodata&   12&           X-ray\\
 11&   15&  \nodata&   20&       X-ray,SMG\\
 12&  \nodata&   40&   31&               OF\\
 13&  \nodata&  \nodata&   35&             SMG\\
 14&   12&  \nodata&    8&      X-ray,LIRG\\
 15&  \nodata&   16&   12&               OF\\
 16&   20&  \nodata&   12&     OF,X-ray,SMG\\
 17&  \nodata&   45&  \nodata&            NL-AGN\\
 18&  \nodata&   28&   20&       X-ray,SMG\\
 19&  \nodata&   28&   20&         OF,X-ray\\
 20&  \nodata&   29&   20&               OF\\
 21&  \nodata&   10&    6&           X-ray\\
 22&   10&  \nodata&   12&             LIRG
\enddata
\tablecomments{$^{a}$ Observations were made using the IRS staring mode which observes each target at two nod positions per cycle.  $^{b}$ Ramp times per cycle.  $^{c}$ Selection criteria: OF - optically faint; X-ray detected; SMG - submillimeter galaxy \citep[included in][]{ap08a};  LIRG - typical LIRG; NL-AGN - narrow-line AGN. }
\end{deluxetable}

\section{Data Analysis} 
In the following section we describe our methods for calculating infrared (IR; $8-1000~\micron$) luminosities ($L_{\rm IR}$) for each of the sample galaxies along with IR and UV (1500~\AA~)  based SFRs.
For all calculations we assume a Hubble constant of 71~km~s$^{-1}$, and a standard $\Lambda$CDM cosmology with $\Omega_{\rm M} = 0.27$ and $\Omega_{\Lambda} = 0.73$.   
Due to the small number of sample galaxies, all average properties of the sample are given as a median rather than taking a mean.  

\subsection{Determining Infrared Luminosities from SED Fitting \label{sec-sed}}
Infrared luminosities are extrapolated by fitting the mid-infrared spectra and photometric data to the SED templates of \citet{ce01}, and then integrating between $8-1000~\micron$.  
We choose to fit to the templates of \citet{ce01} since they have been found to exhibit 24/70~$\micron$ flux density ratios which are more representative of the average actually measured for galaxies at $z\sim1$ than either the \citet{dh02} or \citet{gl03} templates \citep{bm09}.  
The best-fit SEDs are determined by a $\chi^{2}$ minimization procedure in which the SED templates are allowed to scale such that they are being fitted for luminosity and temperature separately.   
Consequently, the amplitude and shape of the SEDs scale independently to best match the observations.    
Weights are derived using the uncertainties of the input photometry (i.e. the S/N ratio of each input photometric band).    
When fitting the mid-infrared spectra, having $N$ individual measurements for each wavelength in the spectra, the 
individual flux density is weighted by its S/N ratio and an additional factor of $1/N$ so as to not give too much weight to the mid-infrared portion of the spectrum in the fitting.    
The normalization constant is then determined by a weighted sum of observed-to-template flux density ratios for all input data used in the fitting.  
Errors in the fits are determined by a standard Monte Carlo approach using the photometric uncertainties of the input flux densities.  

Since we are interested to see to what extent the mid-infrared excesses reported by \citet{ed07a,ed07b} can be 
solely attributed to improper SED fitting rather than the presence of an embedded AGN, we compute IR luminosities in a number of ways.  
First, like \citet{ed07a}, we perform the SED fitting using the 24~$\micron$ flux density alone; \citet{ce01} and \citet{dd01} SED templates are fit independently, and their integrated $8-1000~\micron$ luminosities are then averaged.  
The associated IR luminosity is designated as $L_{\rm IR}^{24}$.  
The difference between the resultant $L_{\rm IR}^{24}$ from the two sets of SED templates is characterized as a systematic error. 
We emphasize here that IR luminosities derived in this manner assume that the mid-infrared to total-infrared correlations imprinted in local SED templates are valid at all redshifts.

Next, we use all available spectroscopic and photometric data in the SED fitting: IRS spectra, 16, 24, 70, and 850~$\micron$ flux densities where possible.  
IR luminosities for 4 out of the 22
sources that are neither detected at 70 or 850~$\micron$ (i.e. GN-IRS~12, 17, 19, and 20) are estimated by averaging the IR luminosities from the best-fit \citet{ce01} and \citet{dd01} SEDs. Due to the non-detection of the 4 sources
at wavelengths close to the peak of the far-infrared emission, these IR luminosities are considered to be upper limits.  
In the cases where only upper limit values are available for the 
photometric data, they are not incorporated into the $\chi^{2}$ minimization but are used to reject fits which have flux densities above the associated upper limit.  
The IR luminosities derived from these fits are labeled as $L_{\rm IR}^{\rm all}$ and are assumed to be the best-fit estimate of the true IR luminosity for each source.  
In addition, we also calculate IR luminosities by fitting the SED templates to all available photometry (i.e. including the 850~$\micron$ data where possible) to assess how well the PAH features can be characterized by fitting photometric data alone.  
These values are labeled as $L_{\rm IR}^{\rm phot}$.  
We give $L_{\rm IR}^{24}$, $L_{\rm IR}^{\rm all}$, and $L_{\rm IR}^{\rm phot}$ values for each galaxy in Table~\ref{tbl-4}

Since it is more likely that mid-infrared spectroscopy and submillimeter imaging will not be available for a large number of sources in many deep fields, we also calculate IR luminosities as above excluding these data.
A comparison of these IR luminosities, identified as $L_{\rm IR}^{16,24,70}$, to those calculated using all available photometry will help quantify the reliability of SED fitting using $16 - 70~\micron$ photometry alone.
The IR luminosities derived by this fitting method are also given in Table~\ref{tbl-4} and are considered to be upper limits for sources not detected at 70~$\micron$.    

To illustrate the differences in the best-fitting SEDs using these four methods, we plot the best-fitting \citet{ce01} templates along with the spectroscopic and photometric data, for 3 different sources (GN-IRS~2, 4, and 15) in Figure \ref{fig-2}.  
A comparison of each fitting method zoomed in on the mid-infrared range of $5-15~\micron$, where PAH emission may be present, is also shown for each galaxy in Figure \ref{fig-3} along with the actual spectra.    
While the SMGs in our sample have been presented in \citet{ap08a}, we note that our SED fitting method differs from theirs. 
We do not include an extinction term and exclude the radio photometry which allows us to assess the FIR-radio correlation among the galaxies.  
Consequently, for the SMGs in our sample we find that the \citet{ap08a} luminosities range from being a factor of $\sim$0.8 to $\sim$4.6 times that of our derived IR luminosities, and are a factor of $\sim$1.2 times larger, on average.      

\subsection{Infrared Luminosities from the Radio \label{sec-firrc}}
In the local Universe there exists a remarkably tight correlation between the far-infrared (FIR; $42.5 - 122.5~\micron$) and predominantly non-thermal radio continuum emission from star-forming galaxies \citep{de85, gxh85,yun01}. 
Following a similar quantitative treatment of the FIR-radio correlation \citep[i.e.][]{gxh85}, we use the total IR ($8-1000~\micron$) luminosity, rather than the FIR fraction, to parameterize the IR-radio correlation such that,    
\begin{equation}
\label{eq-qIR}
q_{\rm IR} \equiv \log~\left(\frac{L_{\rm IR}}{3.75\times10^{12}L_{\nu}(20~{\rm cm})}\right).
\end{equation}
For a sample of 164 galaxies without signs of AGN activity, \citet{efb03} report a median $q_{\rm IR}$  value of 2.64 and a scatter of 0.26~dex. 

We can estimate the rest-frame 1.4~GHz radio luminosities of our spectroscopic sample using, 
\begin{equation}
\label{eq-lrc}
L_{\nu}{(20~{\rm cm})} = 4\pi D_{\rm L}^{2} S_{\nu}(20~{\rm cm})(1+z)^{\alpha-1},
\end{equation} 
where $D_{\rm L}$ is the luminosity distance.  
This calculation includes a bandwidth compression term of $(1+z)^{-1}$ and a $K$-correction of $(1+z)^{\alpha}$ to rest frame 1.4~GHz.   
This assumes a synchrotron power law of the form $S_{\nu} \propto \nu^{-\alpha}$ where the spectral index $\alpha\sim$0.8 \citep{jc92}.  

Thus, by setting  $<q_{\rm IR}> \approx 2.64$, we can derive the infrared luminosities from the radio specific luminosities with:
\begin{equation}
\label{eq-l_IRrc}
L_{\rm IR}^{\rm RC} = 1.64\times10^{15}L_{\nu}(20~{\rm cm})
\end{equation}
We assign a factor of $\sim$2 uncertainty to these luminosities since this is the intrinsic scatter among and within \citep[e.g.][]{ejm06,ejm08} star-forming systems in the local Universe.  
The radio based IR luminosities and $q_{\rm IR}$ ratios are given in Table \ref{tbl-4} for each galaxy.
  
\setcounter{table}{3}
\begin{deluxetable*}{ccccccccc}
\tablecaption{Infrared Luminosities and Associated Parameters\label{tbl-4}}
 \tabletypesize{\scriptsize}
\tablewidth{0pt}
\tablehead{
\colhead{} & \colhead{$\log~L_{\rm IR}^{24}$} & \colhead{$\log~L_{\rm IR}^{16,24,70}$} & 
\colhead{$\log~L_{\rm IR}^{\rm phot}$} & \colhead{$\log~L_{\rm IR}^{\rm all}$} & 
\colhead{$\log~L_{\rm IR}^{\rm RC}$} & \colhead{$\log~L_{\rm AGN}$} & 
\colhead{mid-IR AGN} & \colhead{}\\ 
\colhead{ID} & \colhead{($L_{\sun}$)} & \colhead{($L_{\sun}$)} &
\colhead{($L_{\sun}$)} & \colhead{($L_{\sun}$)} & \colhead{($L_{\sun}$)} &
\colhead{($L_{\sun}$)} & \colhead{Fraction (\%)} & \colhead{$q_{\rm IR}$}\\
\colhead{(1)} & \colhead{(2)} & \colhead{(3)} & \colhead{(4)} & \colhead{(5)} &
\colhead{(6)} & \colhead{(7)} & \colhead{(8)} & \colhead{(9)} 
}
\startdata
  1&  12.79&   $<$12.46&      12.12&      12.12&  13.25&          11.89&       0.50&   1.51\\
  2&  13.62&      12.90&      12.90&      13.04&  13.12&          12.69&       0.79&   2.56\\
  3&  11.85&      11.79&      11.79&      11.78&  12.00&      $<$ 11.04&   $<$ 0.25&   2.42\\
  4&  13.49&   $<$13.00&      12.48&      12.83&  12.79&          12.21&       0.49&   2.69\\
  5&  12.86&   $<$12.44&      12.73&      12.73&  13.25&          11.86&       0.46&   2.13\\
  6&  12.84&   $<$12.60&      12.73&      12.71&  12.45&      $<$ 11.78&   $<$ 0.48&   2.90\\
  7&  12.87&   $<$12.29&      12.21&      12.21&  13.23&          11.43&       0.18&   1.62\\
  8&  11.64&      11.64&      11.64&      11.59&  11.41&      $<$ 10.77&   $<$ 0.23&   2.82\\
  9&  12.77&   $<$12.54&      12.09&      12.09&  12.88&          11.63&       0.30&   1.86\\
 10&  12.03&      11.86&      11.86&      11.89&  11.86&          11.35&       0.44&   2.67\\
 11&  12.28&      12.39&      12.38&      12.38&  12.81&      $<$ 11.63&   $<$ 0.36&   2.22\\
 12&  12.93&   $<$12.48&   $<$12.48&   $<$12.33&  12.93&      $<$ 11.83&   $<$ 0.52&   2.05\\
 13&  11.66&   $<$11.50&      11.49&      11.50&  11.80&      $<$ 11.34&   $<$ 0.51&   2.35\\
 14&  11.90&      11.91&      11.91&      11.91&  12.15&          10.99&       0.22&   2.41\\
 15&  13.55&      12.78&      12.78&      12.76&  12.74&          12.44&       0.69&   2.66\\
 16&  13.07&   $<$12.53&      12.39&      12.39&  12.90&      $<$ 11.93&   $<$ 0.51&   2.14\\
 17&  13.19&   $<$12.51&   $<$12.51&   $<$12.74&  12.56&          12.42&       0.95&   2.82\\
 18&  13.13&   $<$12.54&      12.35&      12.35&  13.11&      $<$ 11.80&   $<$ 0.33&   1.89\\
 19&  12.94&   $<$12.23&   $<$12.24&   $<$12.23&  12.84&          11.69&       0.29&   2.04\\
 20&  12.97&   $<$12.47&   $<$12.47&   $<$12.29&  12.47&          11.86&       0.41&   2.47\\
 21&  13.23&      12.65&      12.65&      12.65&  12.87&          12.40&       0.62&   2.43\\
 22&  11.66&      11.60&      11.60&      11.60&  11.92&      $<$ 11.04&   $<$ 0.29&   2.32
 \enddata
\tablecomments{Col (1): Source ID.  Col (2): 24-$\micron$ derived IR luminosity.  Col (3): IR luminosity derived from SED fitting photometric {\it Spitzer} data: 16, 24, and 70~$\micron$ photometry where available. Col (4): IR luminosity derived from SED fitting all photometric data: 16, 24, 70, and 850~$\micron$ photometry where available. Col (5): Best-fit IR luminosites derived by SED fitting all available data: IRS spectroscopy, 16, 24, 70, and 850~$\micron$ photometry where available.  Col (6): IR luminosity derived using K-corrected 20~cm flux densities and the FIR-radio correlation (see  $\S$\ref{sec-firrc}).  Col (7): The AGN contribution to the total IR luminosity derived using a mid-infrared spectral decomposition of the IRS data (see $\S$\ref{sec-AGNfrac}).  Col (8): The mid-infrared AGN fractions (see $\S$3.5).  Col (9): The logarithmic IR/radio ratio using the usual $q$-based parameterization given in Equation \ref{eq-qIR}.}  
\end{deluxetable*}

\subsection{Estimating Star-formation Rates} 
\subsubsection{IR Based SFRs}
Using the conversion given in \citet{rck98}, calibrated for a \citet{es55} IMF with mass limits between 0.1 and 100~$M_{\sun}$, we can express our IR luminosities as SFRs such that 
\begin{equation}
\label{eq-SFRIR}
	\left(\frac{{\rm SFR_{IR}}}{M_{\sun}~{\rm yr}^{-1}}\right) = 1.73 \times 10^{-10} \left(\frac{L_{\rm IR}}{L_{\sun}}\right).
\end{equation}
Since we have estimated IR luminosities in different manners, we identify each corresponding SFR accordingly; ${\rm SFR_{IR}^{24}}$: SEDs fit by 24~$\micron$ flux densities alone;  ${\rm SFR_{IR}^{all}}$: SEDs fit using all available data (i.e. our best-fit estimates);  ${\rm SFR_{IR}^{RC}}$: using the IR-radio correlation to derive corresponding IR luminosities for which to estimate a SFR.  

\subsubsection{UV Based SFRs}
We also estimate SFRs using rest-frame UV specific luminosities of each galaxy where possible.  
After putting the observed $UBViz$ data into the rest-frame, 1500~\AA~ flux densities were computed by linearly interpolating along the power-law fits between photometric data points landing in the wavelength range between $1250-2600$~\AA.    
Again taking the \citet{rck98} conversion, assuming the same \citet{es55} IMF ($0.1-100~M_{\sun}$), we can express the resultant UV specific luminosities in terms of a SFR such that,
\begin{equation}
\label{eq-SFRUV}
	\left(\frac{{\rm SFR_{UV}}}{M_{\sun}~{\rm yr}^{-1}}\right) = 5.36\times 10^{5} \left(\frac{L_{\rm UV}}{L_{\sun}~{\rm Hz^{-1}}}\right).
\end{equation}
For a total of 11 galaxies (i.e. half of the sample) there are less than two photometric points in the rest-frame wavelength range of $1250-2600$~\AA.  
In these cases the flux upper limits were used to set a corresponding upper limit on each source's rest-frame UV luminosity.  

To estimate the amount of extinction at 1500~\AA~ we use the method described by \citet{gm99} which relates the amount of extinction to the slope of the UV continuum between $1250-2600$~\AA~, defined as $\beta$.  
Since the redshifts of the sample galaxies are all less than $z\sim2.6$, any correction to the UV slope due to absorption by the intergalactic medium is negligible.   
While the relation of \citet{gm99} is calibrated for $A_{1600}$, we use modeled extinction curves \citep[i.e.][]{wd01,btd03}, to translate this into an extinction at 1500~\AA~ such that \(A_{1500} = 1.05 A_{1600}\) leading to a final relation of 
\begin{equation}
\label{eq-extcor}
A_{1500} = 4.65 + 2.09 \beta.
\end{equation}
We designate SFRs estimated using the extinction corrected UV luminosities as  ${\rm SFR_{\rm UV}^{\rm corr}}$.  
Of the 11 sources having less than two photometric detections in the rest-frame wavelength range of $1250-2600$~\AA, two (GN~IRS~16 and 20) have at least one detection and one reliable upper limit in this wavelength range allowing us to estimate a reasonable lower limit for $\beta$.   
For the remaining 9 galaxies, a UV extinction correction was not applied (i.e. $\beta \approx -2.2$).  
The observed and extinction corrected UV specific luminosities are given, along with the values of $\beta$, in Table \ref{tbl-5}.    

For the 11 sources which do not have a sufficient number of photometric points in the $1250-2600$~\AA~range to properly measure $\beta$, the SFR
is considered to be a lower limit given that the extinction correction could not be measured and is likely significant for these sources;  
i.e. due to the fact that the sample is 24~$\micron$ selected, the optical non-detections are probably not from the source being intrinsically dim, but rather from the extinction factor being extremely large.  

\subsection{X-ray Luminosities}  
To compare IR and X-ray energetics of these systems, the rest-frame $0.5-8.0$~keV and $2.0-8.0$ X-ray luminosities are calculated as 
\begin{equation}
\label{eq-lxray}
L_{\rm X} = 4\pi D_{\rm L}^{2} f_{\rm X}(1+z)^{\Gamma-2} 
\end{equation}
where $D_{\rm L}$ is the luminosity distance, $f_{\rm X}$ is the X-ray flux in a particular band, and $\Gamma$ is the observed or assumed photon index (Table~\ref{tbl-3}).  
The $0.5-8.0$ and $2.0-8.0$~keV luminosities are given in Table~\ref{tbl-5}.  
We also note here that 3 of the 12 X-ray sources (i.e. 25\%) are optically faint, having only IRS-derived redshifts between $1.7 \la z \la 1.9$.   
This is precisely the redshift range where optically faint X-ray sources are predicted to lie \citep{da01,vm05}.

\subsection{Estimating AGN Contributions \label{sec-AGNfrac}}
Using the IRS data, we follow the procedure described in detail by \cite{ap08a} to quantify the AGN contribution to the total IR energy budget.
This is done by first decomposing the IRS spectra into two components; one associated with star formation (PAH template) and the other associated with the hot dust continuum emission arising from an embedded AGN.  
The star-forming component is fit using the starburst composite template of \citet{bb06} while the AGN component is characterized by a power law with both normalization and slope as free parameters.  
The effect of extinction is considered independently for the PAH and continuum components using a modeled extinction curve \citep[i.e][]{wd01,btd03}.  
Upon finding the best fit between the spectra and combination of PAH template plus continuum component, the fraction of mid-infrared emission arising from the continuum component is taken to be the AGN contribution.
For those galaxies whose spectra did not have high enough S/N ratios or lacked PAH features to allow for a 
proper decomposition, an upper limit to the AGN strength was obtained by fitting a power-law 
to the continuum of the spectra and integrating below it.  
The mid-infrared AGN fractions are given for each galaxy in Table~\ref{tbl-4}.  

\begin{figure}
\plotone{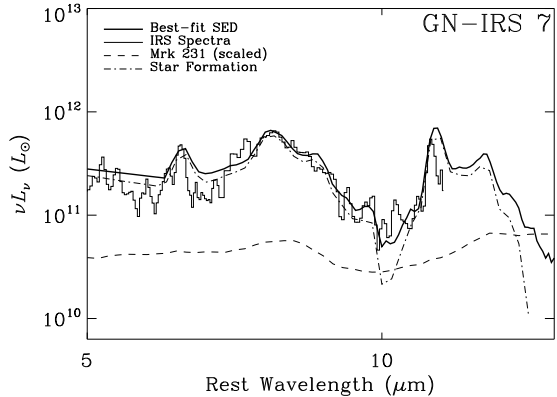}
\caption{
The best-fit SEDs of \citet{ce01} for GN-IRS~7 plotted as the bold {\it solid} line, given in the rest-frame.  
Overplotted are the IRS spectra of the source ({\it solid histogram-style} line), the scaled Mrk~231 template ({\it dashed} line), corresponding to the contribution to the SED by the AGN, and the contribution to the SED by star formation ({\it dot-dashed} line).  
\label{fig-4}}
\end{figure}

Next, to characterize the fraction of the IR luminosity arising from an AGN, we scale the SED template of Mrk~231 \citep[a nearby ULIRG whose IR luminosity is known to be dominated by an AGN;][]{la07} to the AGN fraction of the rest-frame $\sim$$5-12~\micron$ luminosity (i.e. the luminosity in the rest-frame wavelength range having IRS coverage per source) and integrate the SED from $8-1000~\micron$.  
This is illustrated in Figure \ref{fig-4} where we show the best-fit \citet{ce01} SED for GN-IRS~7 along with the observed IRS spectra, the scaled SED of Mrk~231 (i.e. the AGN contribution), and the difference between the best-fit SED and the scaled SED of Mrk~231 (i.e. the contribution by star formation alone).  
The observed mid-infrared spectra clearly matches the shape of the best-fit SED.  

The values of $L_{\rm IR}^{\rm AGN}$ are given for each galaxy in Table \ref{tbl-4}.  
By subtracting these values from the IR luminosites obtained through our SED fitting, we can estimate the amount of IR luminosity arising from star formation alone such that
\begin{equation}
\left(\frac{\rm SFR_{\rm IR}^{\rm noAGN}}{M_{\sun}~{\rm yr}^{-1}}\right) = 
	1.73\times10^{-10}\left(\frac{L_{\rm IR}^{\rm all,noAGN}} {L_{\sun}}\right),
\end{equation}
where $L_{\rm IR}^{\rm noAGN} = L_{\rm IR}^{\rm all} - L_{\rm IR}^{\rm AGN}$.    
Similarly, we can also calculate an AGN-subtracted SFR from the 24~$\micron$ data by SED fitting the 24~$\micron$ photometry after first subtracting off  
AGN fractions extrapolated from those given in Table \ref{tbl-4}; 
the corresponding IR luminosities and SFRs are denoted as $L_{\rm IR}^{\rm24,noAGN}$ and ${\rm SFR}_{\rm IR}^{\rm24,noAGN}$, respectively.    
We have assumed that the underlying mid-infrared continuum arises purely from an AGN, however hot dust associated with vigorous star formation may also contribute to the mid-infrared continuum emission as well.  
Therefore, the mid-infrared AGN fractions, and associated estimates for the IR luminosities of the AGN, should all be considered conservative upper limits.

\begin{deluxetable*}{cccccc}
\tablecaption{X-ray and UV Luminosities\label{tbl-5}}
 \tabletypesize{\scriptsize}
\tablewidth{0pt}
\tablehead{
\colhead{} & 
\colhead{$\log~L_{\rm 0.5-8.0~keV}$} & \colhead{$\log~L_{\rm 2.0-8.0~keV}$} &
\colhead{$\log~L_{\rm UV}$} & \colhead{$\log~L_{\rm UV}^{\rm corr}$} & \colhead{}\\ 
\colhead{ID} & \colhead{($L_{\sun}$)} & \colhead{($L_{\sun}$)} &
\colhead{($L_{\sun}$~Hz$^{-1}$)} & \colhead{($L_{\sun}$~Hz$^{-1}$)} & \colhead{$\beta$}\\
\colhead{(1)} & \colhead{(2)} & \colhead{(3)} & \colhead{(4)} & \colhead{(5)} 
}
\startdata
  1&      10.16&      10.13&   $<$-5.50&   $<$-5.50&        \nodata\\
  2&   $<$ 8.91&   $<$ 9.14&      -4.74&      -3.23&      -0.47\\
  3&       8.14&   $<$ 8.03&   $<$-4.94&   $<$-4.94&        \nodata\\
  4&       9.59&       9.49&      -5.11&      -3.09&       0.14\\
  5&   $<$ 8.75&   $<$ 8.94&   $<$-5.43&   $<$-5.43&        \nodata\\
  6&   $<$ 8.85&   $<$ 9.07&   $<$-5.32&   $<$-5.32&        \nodata\\
  7&   $<$ 8.64&   $<$ 8.84&   $<$-5.44&   $<$-5.44&        \nodata\\
  8&       7.96&       7.86&   $<$-4.78&   $<$-4.78&        \nodata\\
  9&       8.67&       8.72&      -4.85&      -3.46&      -0.61\\
 10&       9.79&       9.48&   $<$-4.41&   $<$-4.41&        \nodata\\
 11&       8.69&   $<$ 8.47&      -4.83&      -2.96&      -0.03\\
 12&   $<$ 8.81&   $<$ 9.04&      -4.98&      -3.76&      -0.82\\
 13&   $<$ 7.85&   $<$ 8.05&      -5.20&      -3.18&       0.14\\
 14&       8.12&   $<$ 8.16&      -6.18&      -3.69&       0.71\\
 15&   $<$ 8.98&   $<$ 9.24&      -5.18&      -4.04&      -0.91\\
 16&       8.43&   $<$ 8.56&   $<$-5.59&   $<$-4.04&   $<$-0.42\\
 17&   $<$ 8.80&   $<$ 9.05&      -4.23&      -3.96&      -1.95\\
 18&       8.97&       8.94&      -4.57&      -2.53&       0.18\\
 19&       8.75&       8.73&   $<$-5.53&   $<$-5.53&        \nodata\\
 20&   $<$ 9.24&   $<$ 9.49&   $<$-5.50&   $<$-2.69&   $<$ 1.09\\
 21&       9.23&       9.20&      -5.55&      -2.48&       1.40\\
 22&   $<$ 7.87&   $<$ 8.07&   $<$-5.60&   $<$-5.60&        \nodata
 \enddata
\tablecomments{Col (1): Source ID.  Col (2): K-corrected full band ($0.5-8.0$~keV) luminosities.  Col (3): K-corrected hard band ($2.0-8.0$~keV) luminosities.   Col (4): Linearly interpolated 1500~\AA~ specific luminosities. Col(5): Extinction corrected 1500~\AA~ specific luminosities.  Col(6): Slope of the UV continuum between $1250-2600$~\AA~ which is used to estimate extinction following \citet{gm99}.  }  
\end{deluxetable*}

\section{Results}
We now present the results for the various SED fitting methods among the sample while looking for trends with redshift.      
In doing so, we also compare the infrared-based SFRs to those estimated from extinction corrected UV emission for galaxies having redshifts between $1.4\la z \la2.6$.  

\subsection{Accuracy of Derived IR Luminosities \label{sec-fitcomp}}
\begin{figure}
\plotone{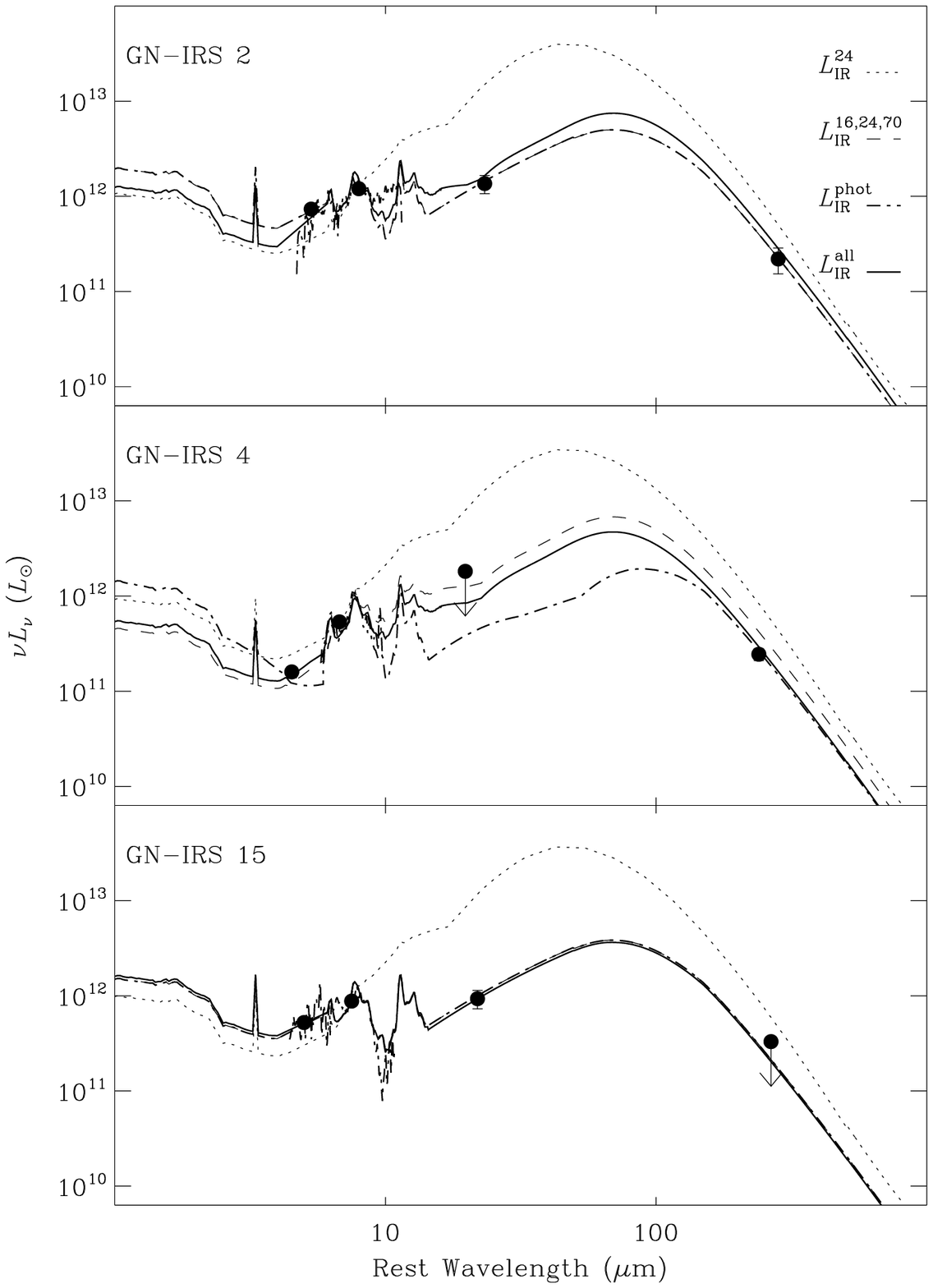}
\caption{
The best-fit SEDs of \citet{ce01} for GN-IRS~2 (top panel), 4 (middle panel), and 15 (bottom panel) using only 24~$\micron$ flux densities ($\it dotted$ lines), all 3 {\it Spitzer} (16, 24, and 70~$\micron$) flux densities ({\it dashed} lines), all {\it Spitzer} and 850~$\micron$ photometry ({\it dot-dashed} lines), and all {\it Spitzer} photometry along with the IRS spectra and the 850~$\micron$ flux densities  ({\it solid} lines).  
The best-fit SEDs associated with $L_{\rm IR}^{16,24,70}$, $L_{\rm IR}^{\rm phot}$, and $L_{\rm IR}^{\rm all}$ for GN-IRS~15 are very similar, which is why the {\it dashed} and {\it dot-dashed} lines overlay each other. 
\label{fig-2}}
\end{figure}

\begin{figure*}
\plotone{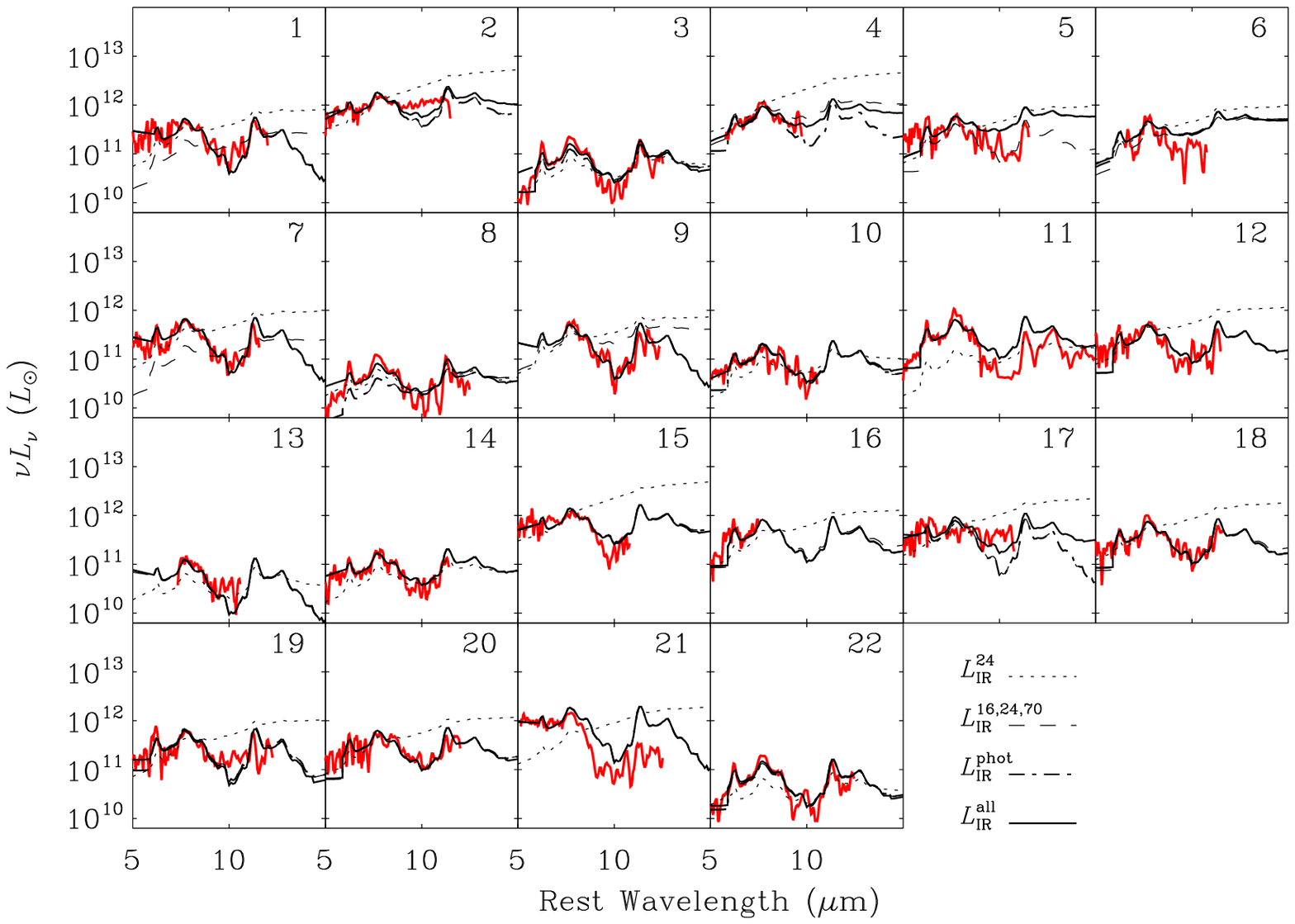}
\caption{
A blowup of the $5-15~\micron$ range for the best-fit SEDs of \citet{ce01} for all sources using only 24~$\micron$ flux densities ($\it dotted$ lines), all 3 {\it Spitzer} (16, 24, and 70~$\micron$) flux densities ({\it dashed} lines), {\it Spitzer} and submillimeter (850~$\micron$) photometry ({dot-dashed} lines), and all photometry ({\it Spitzer} and submillimeter) along with the IRS spectra ({\it solid} lines).  
The actual spectra, smoothed to the instrumental resolution, are overplotted in red.  
The SEDs chosen when fitting with the 24~$\micron$ data alone rarely characterize the observed PAH emission as 
compared to when longer wavelength data is included in the fitting.    
\label{fig-3}}
\end{figure*}

In Figure \ref{fig-2} we plot the best-fitting \citet{ce01} SEDs for 3 sources (GN-IRS~2, 4, and 15) to illustrate the three fitting methods.  
While the  addition of the IRS spectra and submillimeter data to the available  {\it Spitzer} 16, 24, and 70~$\micron$ photometry does not seem to significantly alter the choice of the fit, the fits using 24~$\micron$ flux densities alone are highly discrepant.  
Zooming in on the $5-15~\micron$ mid-infrared regions of the four different \citet{ce01} fits for each galaxy in Figure \ref{fig-3}, one easily sees that by using 24~$\micron$ flux densities alone the equivalent widths of the aromatic features observed in the IRS spectra are 
underestimated by the \citet{ce01} templates, resulting in highly discrepant estimates of IR luminosity.  

The range of IR luminosities derived by fitting the 24~$\micron$ photometry alone span a range between $4.3\times10^{11}-4.2\times10^{13}~L_{\sun}$ (i.e. nearly two orders of magnitude) with a median of $7.5\times10^{12}~L_{\sun}$.  
Looking at the best-fit estimates for the IR luminosities among the entire sample, we find that they range between $3.2\times10^{11}-1.1\times10^{13}~L_{\sun}$ (i.e. a factor of $\sim$35) having a median of $2.2\times10^{12}~L_{\sun}$.  
The 24~$\micron$-derived IR luminosity is a factor of $\sim$4 larger than that for the best-fit estimates, on average.  

Using our best-fit IR luminosites, we also note that 16 out of the 22 sample galaxies are classified as ULIRGs while the remaining 6 are LIRGs.  
Looking at the differences between the 24~$\micron$ to our best-fit IR luminosities, the discrepancy is found to be much larger among the ULIRGs than for the LIRGs.  
The 24~$\micron$-derived IR luminosities are larger than the best-fit IR luminosities by a factor of $\sim$5, on average, for the ULIRGs while being only $\sim$1.2 times larger, on average, for the LIRGs.  

\subsection{Strength of PAH Features}  
To quantify how well the mid-infrared spectral features are matched by the various SED fitting methods, we compare the ratio from the peak of the 7.7~$\micron$ PAH feature to the median flux density in the $9-11~\micron$ trough region (i.e. 9.7~$\micron$ silicate absorption feature).    
Given that the wavelength coverage varies for each object, making it difficult to properly estimate the true continuum levels \citep[e.g.][]{ap08a}, we choose to compare this ratio rather than actual PAH equivalent widths.  
Furthermore, choosing these features enables us to calculate this ratio for nearly the entire sample; we do not have spectroscopic coverage over the rest-frame $9-11~\micron$ region for GN-IRS~16.  

When fitting the SEDs using the 24~$\micron$ photometry alone the peak-to-trough ratios are $\sim$$0.24\pm0.15$ times the peak-to-trough ratios measured for the observed spectra (i.e. a factor of $\sim$4 times smaller), on average.   
By instead fitting the SEDs using all of the photometric data, the peak-to-trough ratios are now $\sim$$0.75\pm0.84$ times the ratios of the actual data (i.e. $\sim$35\% smaller), on average.  
Similarly, when the mid-infrared spectra is included in the fitting with all of the photometric data, the peak-to-trough ratios are found to be $\sim$$0.75\pm0.58$ times the ratios for the the observed spectra, on average, suggesting that fitting with mid- and far-infrared photometry alone does a reasonable job choosing templates with the correct mid-infrared spectral features.  
Consequently, it appears that the improper estimates of IR luminosities  from fitting with 24~$\micron$ photometry alone are most likely due to the fact that local high luminosity galaxies do not exhibit large PAH equivalent widths relative to galaxies of similar luminosity at higher redshifts.  
This result is consistent with that of \citet{vd07} who showed that local ULIRGs typically have PAH equivalent widths that are significantly smaller than those measured for ULIRGs at high-redshift.    
Since the strongest PAH features, lying in between 6-12~$\micron$, begin to redshift into the observed 24~$\micron$ band at $z\ga1.4$, we expect the discrepancies between the fitting methods and estimates for the IR luminosity of galaxies,  to increase with increasing redshift.
For the sample galaxies in the redshift range between $1.4\la z \la 2.6$, we find that the peak-to-trough ratios of the SEDs chosen by fitting the 24~$\micron$ data alone  and using all of the available photometric data are an average factor of $\sim$$0.19\pm0.11$ (i.e. $\sim$5 times smaller) and $\sim$$0.88\pm0.94$ (i.e. $\sim$14\% smaller) times those measured for the data, respectively.   

We also note that by remeasuring the peak-to-trough ratios of the best-fit SEDs after subtracting the fitted AGN component, the median ratios are $\sim$$1.11\pm0.30$ and $\sim$$1.49\pm0.42$ for the entire sample and galaxies at $z \ga 1.4$, respectively.   
This result demonstrates that the AGN/star-formation decomposition is fairly reliable and that, as previously stated, we likely overestimate the true AGN contribution by assuming that all of the mid-infrared continuum emission arises from hot dust associated with the AGN.  
These statistics exclude 3 galaxies (GN-IRS~1, 13, 17) for which the AGN component over-subtracts the continuum of the best-fit SEDs resulting in a negative peak-to-trough ratio.  
Including them into the above calculations does not affect the median values, but increases the dispersion significantly.     

\subsection{Redshift Dependency}
\begin{figure}
\plotone{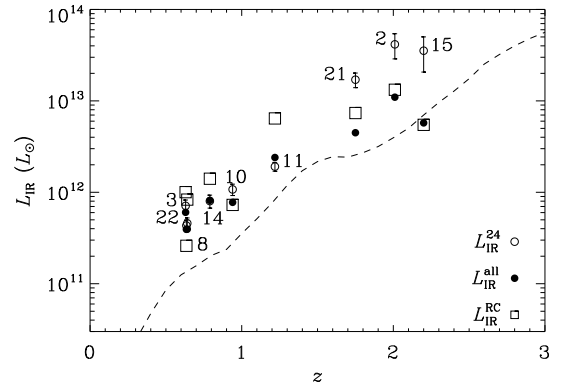}
\caption{
We plot IR luminosity, estimated by three different means, versus redshift for galaxies in our IRS sample which have 70~$\micron$ detections; associated 1-$\sigma$ error bars are given.
$L_{\rm IR}^{24}$ values ({\it open circles}) are calculated by SED template fitting using the 24~$\micron$ flux density only while $L_{\rm IR}^{\rm all}$ values ({\it filled circles}) also include the 16, 70, and 850~$\micron$ photometry (where available) along with IRS spectra.     
$L_{\rm IR}^{\rm RC}$ values are calculated using the well-known FIR-radio correlation and plotted as {\it open squares}; since the uncertainties associated with these estimates are large (i.e. a factor of $\sim$2), we do not plot error bars.
The $L_{\rm IR}^{24}$ estimated by SED fitting the sample's $f_{\nu}(24~\micron)$ flux density limit of 220~$\mu$Jy as a function of redshift is plotted as the {\it dashed} line.    
\label{fig-5}}
\end{figure}

In Figure \ref{fig-5} we plot  $L_{\rm IR}^{24}$, $L_{\rm IR}^{\rm all}$, and $L_{\rm IR}^{\rm RC}$ versus redshift for the 70~$\micron$ detected galaxies.  
Although the different IR luminosity estimates largely agree for the $z<1.4$ sources, the 3 sources at $z>1.4$ have
IR luminosities derived from fitting the 24~$\micron$ flux densities alone that are larger by an average factor of $\sim$4 than what is found when fitting the SEDs with all available data.  

\begin{figure}
\plotone{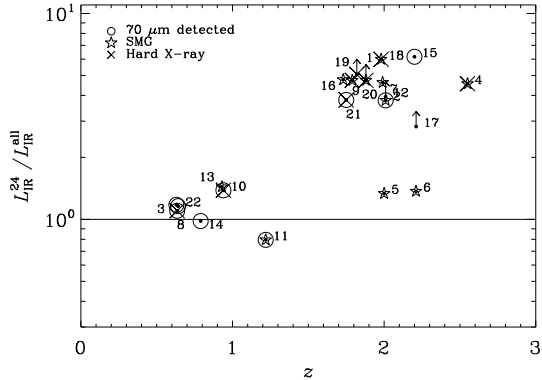}
\caption{
We plot the ratio between the 24~$\micron$-derived IR luminosity ($L_{\rm IR}^{24}$) to that derived using the additional constraints from the IRS spectroscopy along with the 16, 70, and 850~$\micron$ data.  
Sources having hard band ($2.0-8.0$~keV) X-ray detections are identified with an ``X".  
For sources not identified at 70 and 850~$\micron$, we plot an {\it upward arrow} indicating that the IR luminosity is only an upper limit.  
\label{fig-6}}
\end{figure}

Since the 70~$\micron$ detections could potentially be sampling objects with the warmest FIR dust color temperatures, we also
compare the different IR luminosity estimates of all sources in our sample in Figure \ref{fig-6}.  
We find a significant change in this ratio with redshift.
While the objects near $z\sim1$ all have ratios around unity, galaxies in a redshift range between $1.4\la z \la 2.6$ exhibit ratios of $\sim$$4.6\pm1.4$, on average.  
We also note that galaxies lying at redshifts below and above $z \sim 1.4$ typically have $L_{\rm IR}^{\rm 24}$ values below and above $\sim$$3\times10^{12}~L_{\sun}$, respectively.  
This result is consistent with that of \citet{cp07} whom report that IR luminosities derived from stacking 70 and 160~$\micron$ data for a sample of 24~$\micron$-bright (i.e. $f_{\nu}(24~\micron) > 250~\mu$Jy) galaxies lying in a redshift range between $1.5\la z \la 2.5$, are a factor of $2-10$ lower than those inferred from using only 24~$\micron$ flux densities.     
A similar comparison with the radio continuum based luminosities over the same redshift range shows that the 24~$\micron$ based IR luminosities are higher by a factor of $\sim$$1.5\pm1.9$, on average.   

It is unclear if this discrepancy is due to the flux limit of our sample which implies that we are sampling sources with high
luminosities at $z\sim2$ for which no local analogs exist. 
This would suggest that the extrapolation of the local SEDs to these high redshift sources is invalid.  
A second possibility is that $z\sim2$ sources have sufficiently low metallicities that the infrared SEDs deviate from those of $z\sim1$ or local galaxies.
A third possibility that has been suggested is that warm dust continuum emission from an obscured AGN in these sources is biasing the $24\,\mu$m flux upward and resulting in overestimates of the SFR \citep{ed07a}. 

\subsection{AGN-Subtracted SFR Estimates from 24~$\micron$}
The SFRs inferred from the 24~$\micron$-derived IR luminosities overestimate those inferred from our best-fit IR luminosities, corrected for any contributions from AGN (SFR$_{\rm IR}^{\rm noAGN}$; see $\S$3.5) by a factor of $\sim$6, on average, among the entire sample.   
To see whether this overestimation by fitting the 24~$\micron$ photometry is the result of excess mid-infrared emission arising from hot-dust emission associated with the presence of an AGN, we refit the 24~$\micron$ flux densities to the SED templates after correcting them by the corresponding mid-infrared AGN fraction.  
We find that SFRs calculated using IR luminosities derived by SED fitting 24~$\micron$ flux densities after subtracting the mid-infrared AGN contributions  
are $\sim$$2.5\pm 2.3$ times larger, on average, than the values of SFR$_{\rm IR}^{\rm noAGN}$ among the entire sample.  
There also appears to be a redshift dependence such that the  SFRs calculated using IR luminosities derived by SED fitting 24~$\micron$ flux densities after subtracting the mid-infrared AGN contributions are $\sim$$1.0\pm0.4$ and $\sim$$3.3\pm2.2$ times larger, on average, than the values of  SFR$_{\rm IR}^{\rm noAGN}$ for galaxies at redshifts below and above $z\sim1.4$, respectively.  

We believe that IR luminosities are overestimated by fitting local SEDs to 24~$\micron$ photometry alone because of differences in PAH equivalent widths among galaxies of similar luminosity at different redshifts.   
This conclusion is supported by our findings in $\S$4.2 where we compared the observed PAH strengths to those from the different SED fits.  
Alternatively, this result could also imply that using Mrk~231 to approximate the AGN component is inappropriate. 
However, given that we believe our mid-infrared AGN fractions and IR luminosity estimates for the AGN to be conservative upper limits, this is unlikely to be a dominant effect.      

\subsection{Multiwavelength Comparison between SFR Estimates}  
In order to assess the reliability of the different SFR indicators, we make a cross comparison between the UV, radio and IR SFR estimates.  
We examine how these estimates compare for galaxies located at redshifts below and above $z\sim1.4$.   

\subsubsection{$z<1.4$ Galaxies}
For the 7 galaxies located at $z < 1.4$, we find that SFRs inferred from their 24~$\micron$ data alone range from being $\sim$0.8 to $\sim$1.4 times that of  those SFRs calculated using our best-fit IR luminosities, and are only 16\% larger, on average.  
However, using the radio data to infer a SFR via the FIR-radio correlation, we find that the radio SFRs overestimate those calculated using the best-fit IR luminosities by a factor of $\sim$1.7, on average, and range from being a factor of $\sim$0.7 to $\sim$2.7 times that of the SFRs measured using are best-fit IR luminosities.    
Among these 7 galaxies, only 3 have measurable UV slopes allowing us to calculate a UV corrected SFR.   
These SFRs range from being a factor of $\sim$0.8 to $\sim$6.5 times that of the SFRs measured using our best-fit IR luminosities.  

\subsubsection{$z>1.4$ Galaxies \label{sec-sfr-hiz}}

\begin{figure*}
\plotone{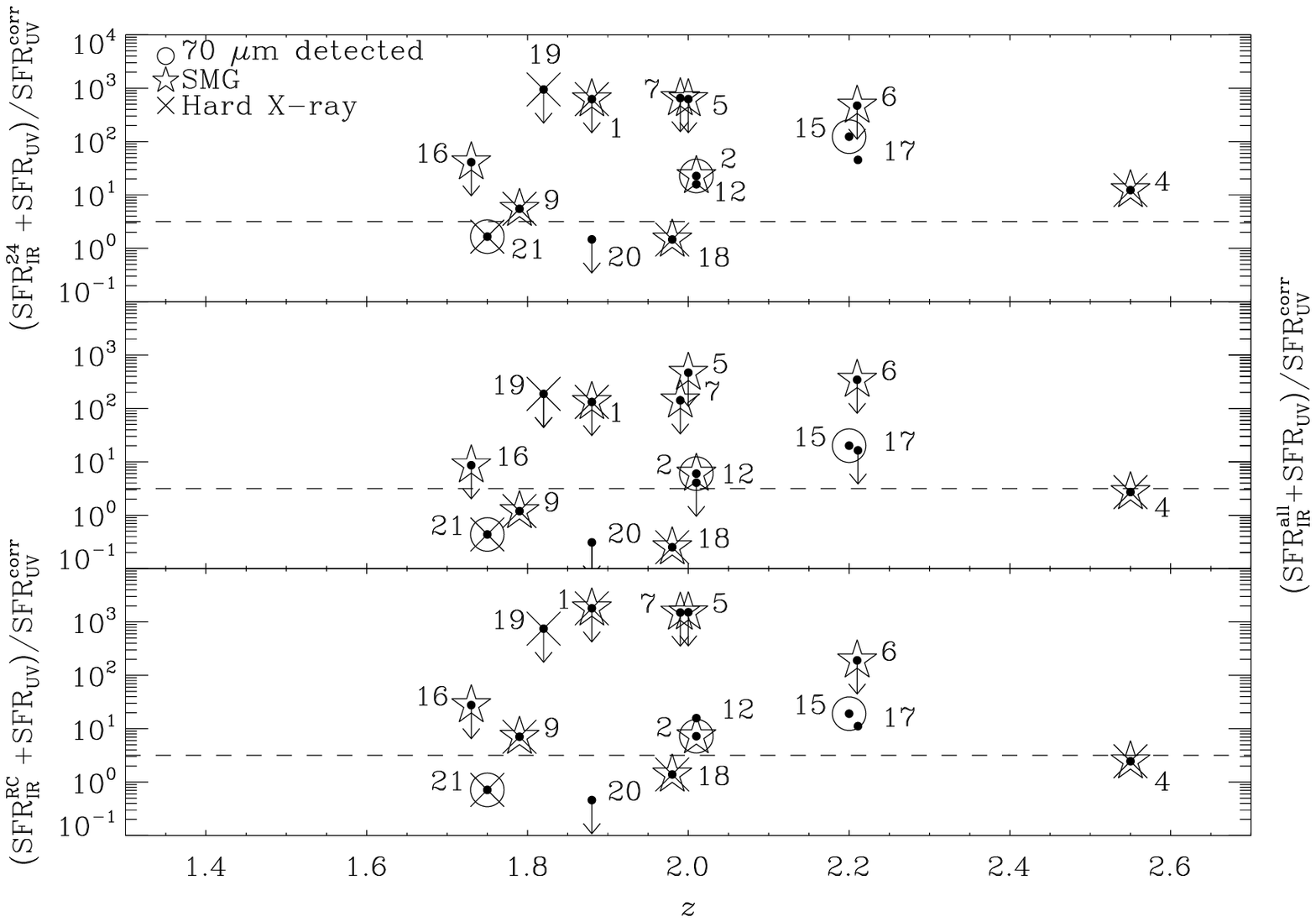}
\caption{
In the top panel we plot the ratio of SFRs estimated from the 24~$\micron$ flux densities (plus a non-obscured UV-derived SFR) to extinction corrected UV SFRs as a function of redshift.   
The {\it dashed} line indicates where the logarithm of this ratio is equal to 0.5~dex; galaxies taking  values larger than this are considered to be ``mid-infrared excess" sources \citep[i.e.][]{ed07a}.   
Sources undetected in the optical, such that its UV slope could not be fit, are plotted with {\it downward arrows}.
In the middle panel we plot the same ratio versus redshift, except that we use all available data to derive the IR based SFRs (${\rm SFR_{IR}^{all}}$). 
The IR based SFRs are considered to be upper limits for those galaxies not detected at 70 and 850~$\micron$
The same is done again in the bottom panel, except that we use the 20~cm radio continuum flux densities and the FIR-radio correlation to derive SFRs (${\rm SFR_{IR}^{RC}}$).
Mid-infrared excess sources largely remain even when the bolometric corrections are accounted for.  
\label{fig-7}}
\end{figure*}

\citet{ed07a} have identified discrepancies between the UV, radio, and, IR derived SFR estimates in $z\sim2$ galaxies selected using the $BzK$ selection technique.
We select the 15 galaxies in our sample with $1.4\la z\la2.6$ which matches the redshift range spanned by the $BzK$ selection.
Figure \ref{fig-7} shows the
the ratio of SFRs inferred from the 24~$\micron$ data (plus the uncorrected UV-based SFR) to those estimated from extinction corrected UV 
measurements for these 15 galaxies. 
Galaxies which are not detected in the optical, such that their UV slopes could not be estimated, are plotted as upper limits
although they are difficult to interpret.    
Their best-fit IR-derived SFRs are nearly 2 orders of magnitude larger than the uncorrected UV-based SFRs, on average.  
However, their extinction corrected UV SFRs could potentially be very large, reducing the plotted ratio.

\citet{ed07a} defined sources for which 
\begin{equation}  
\label{eq-irexc}
\log \left({\rm \frac{SFR_{IR} + SFR_{UV}}{SFR_{UV}^{corr}}}\right) > 0.5 
\end{equation}
to be ``mid-infrared excess" sources.
While the numerator adds the contribution of energetic emission from young massive stars that is not absorbed and reemitted by dust grains to obtain a total SFR, the extincted UV SFR is often negligible compared to the IR-based SFR term.

Not surprisingly, we find that all 7 galaxies for which we do not have UV slope measurements, and thus cannot make UV extinction corrections, all meet the mid-infrared excess criterion.  
For these sources to fall below the mid-infrared excess criterion would require an average extinction of $A_{V}\ga1.7$~mag; and to fall to a ratio near unity requires an average extinction of $A_{V}\ga2.2$~mag.  
As for the remaining 8 sources, we find 6 are identified as having mid-infrared excesses: GN-IRS~2, 4, 9, 12, 15, adn 17.  

The two sources for which we do not find any mid-infrared excess, due to their steep UV corrections, are GN-IRS~18 and 21.  
Both GN-IRS~18 and 21 are X-ray detected, however GN-IRS~18 is also an SMG.  
Morphologically, GN-IRS~18 appears extended possibly due to a major merger event.  
The extinction estimated for GN-IRS~18 is $A_{V} \approx1.92$~mag, only 0.16~mag larger than the median extinction found for all sample galaxies for which the rest-frame UV slope could be measured.  
The extinction estimated for GN~21, however, is the largest among the entire sample at $A_{V} \approx 2.89$~mag.   
While their optical extinctions differ significantly, these are the only two sources having extinction corrected UV SFRs in excess of 1000~$M_{\sun}~{\rm yr}^{-1}$.   

By plotting again the ratio of ${\rm IR + UV}$ to UV corrected SFRs, this time instead using our best-fit IR 
luminosities, versus redshift in the middle panel of Figure \ref{fig-7}, we find that 2 of the 6 previously identified mid-infrared excess galaxies (GN-IRS 4 and 9) are no longer mid-infrared excess sources. 
These two galaxies are also the only ones among the 6 to have hard X-ray detections.  
In fact, only 2 sources, GN-IRS~2 and 15, seem to remain definitive ``mid-infrared excess'' galaxies given that GN-IRS~12, which barely lies above the mid-infrared excess threshold, and GN-IRS~17 are not detected at either 70 or 850~$\micron$ implying that their associated best-fit IR luminosities are technically upper limits.    
If, on the other hand, we use the radio continuum based SFRs to construct the same plot, as shown in the bottom panel of Figure \ref{fig-7}, we find that while points have shifted downward, only GN-IRS~4 is no longer considered a mid-infrared excess source leaving 5 of the original 6 mid-infrared excess galaxies above the threshold.  

To summarize, we find that by using the best-fit IR luminosities we can definitively account for 2/6 mid-infrared excess sources (i.e. $\onethird$ of those found using 24~$\micron$ derived IR luminosities).   
This fraction is increased to $\twothirds$ by assuming that we are still overestimating the IR luminosities of the two galaxies for which we only have upper limits at far-infrared wavelengths.        
The median ratio of best-fit IR to extinction corrected UV SFRs for sources in the redshift range between $1.4\la z\la 2.6$ is $\sim$4.  
By instead using the radio continuum and FIR-radio correlation to derive IR luminosities, the median ratio between the radio derived-to-extinction corrected UV SFRs is even larger at $\sim$7.  
We investigate in the discussion below whether the contribution from an obscured AGN is large enough to account for the discrepancy between the IR and UV SFR estimates among these sources.   


\section{Discussion}
Using IRS spectroscopy along with additional photometric data, we have set out to characterize better the IR properties for a diverse sample of 24~$\micron$ selected sources.  
In doing this we also try to explain the true nature of sources considered to have an excess of IR emission relative to their ultraviolet luminosity when 24~$\micron$ photometry, alone, is used to estimate SFRs. 
Specifically, through a spectral decomposition of the IRS data, we determine that half of the "mid-infrared excess" sources are Compton-thick AGN.  
A summary of our findings for each of the 22 sample galaxies is given in Table \ref{tbl-6}.  

\subsection{Discrepancies in SFR Estimates versus the Contribution from AGN}
As discussed in $\S$~\ref{sec-sfr-hiz}, we find that even after using the best-fit IR luminosities, $L_{\rm IR}^{\rm all}$, to derive SFRs, we still find 4/8 sources in the redshift range between $1.4\la z \la 2.6$ with measurable UV slopes to exhibit excesses in the infrared relative to what is measured by extinction corrected UV data (middle panel of Figure \ref{fig-7}).
To see if the observed excess is in fact due to the presence of an obscured AGN, we estimate the fraction of AGN power contributing to the total IR luminosity using our mid-infrared spectroscopy as described in $\S$\ref{sec-AGNfrac}.
  
\begin{figure*}
\plottwo{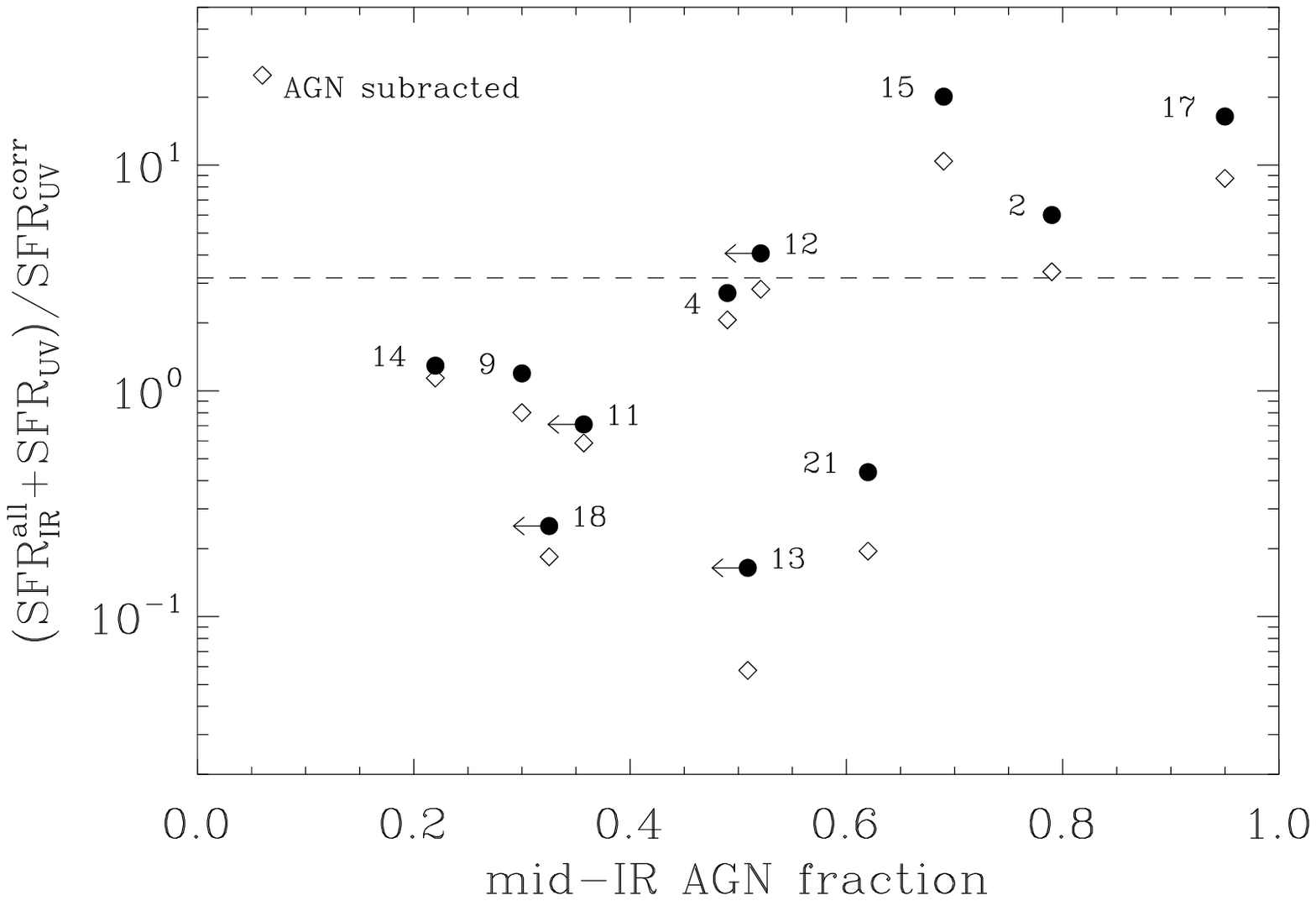}{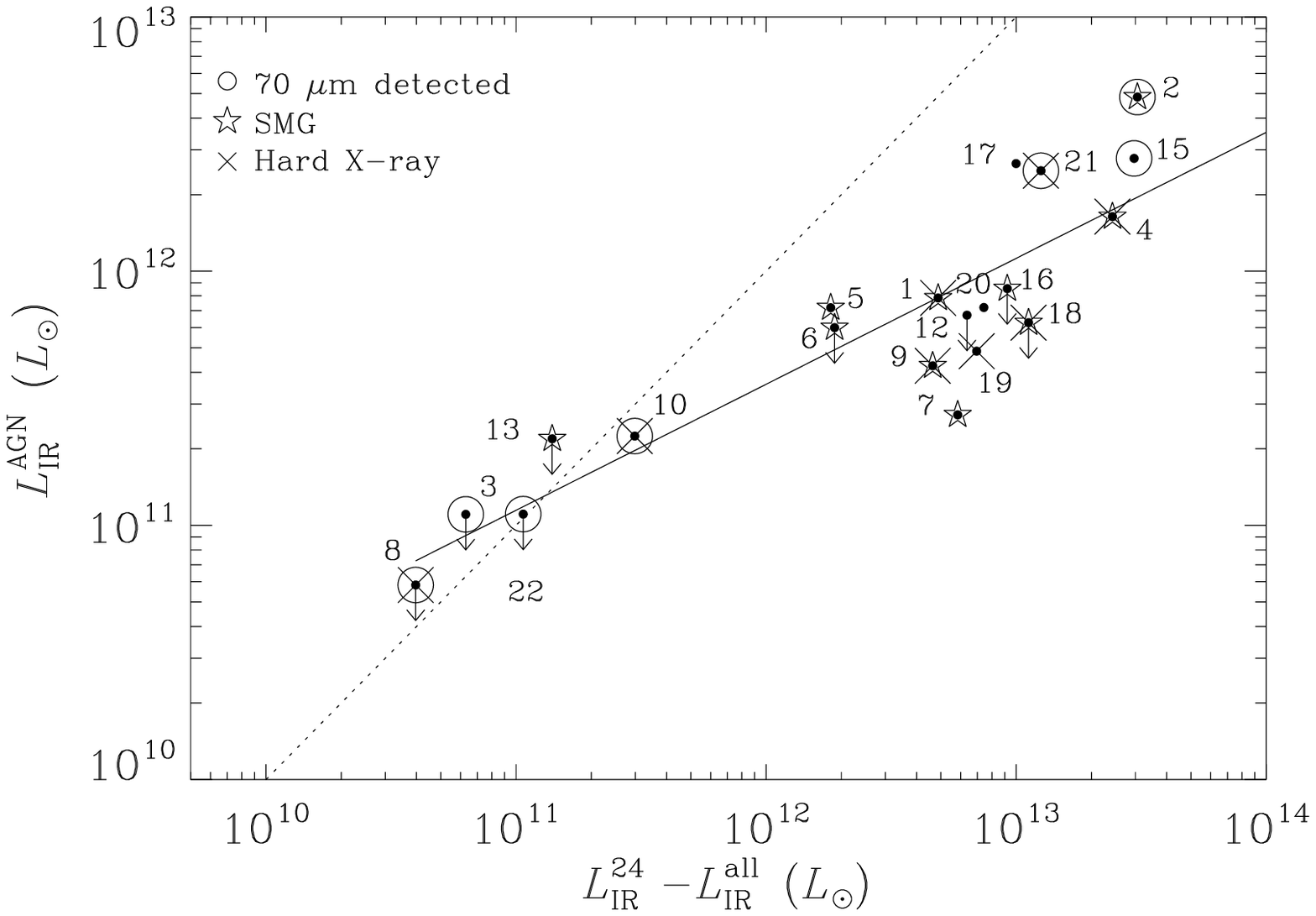}
\caption{
In the left panel we plot the ratio of IR-derived SFRs using our best-fit IR luminosities, plus an unobscured UV component, to extinction corrected UV SFRs versus the mid-infrared AGN fraction before and after subtracting the AGN contribution to the total IR luminosities using {\it filled circles} and {\it open diamonds}, respectively.  
The {\it dashed} line indicates where the logarithm of this ratio is equal to 0.5~dex; galaxies taking  values larger than this are considered to be ``mid-infrared excess" sources \citep[i.e.][]{ed07a}.   
In the right panel we plot the estimated AGN luminosity (see $\S$\ref{sec-AGNfrac}) as a function of the difference between the 24~$\micron$ and best-fit IR luminosities.
Sources for which only upper limits to the AGN mid-infrared luminosity fraction could be determined are shown as {\it downward arrows}.  
An ordinary least squares fit is overplotted as a {\it solid} line while the {\it dotted} line indicates a one-to-one relationship to show that the AGN luminosity is not able to account for the difference in the bolometric correction when fitting the SED templates with 24~$\micron$ photometry alone.     
\label{fig-8}}
\end{figure*}

In the left panel of Figure \ref{fig-8} we plot the ratio of our best-fit IR derived SFRs, plus an unobscured UV component, to the extinction corrected UV SFRs against the mid-infrared AGN fractions before and after subtracting the AGN contribution to the total IR luminosity. 
We find that 3/8 galaxies having redshifts between $1.4\la z \la 2.6$ and which continue to exhibit an excess of IR emission have mid-infrared AGN fractions that are larger than 60\%.  
Using the Mrk 231 SED as outlined in $\S$ 3.5, we can estimate the contribution of AGN to the total IR luminosity.

Looking at the entire sample (i.e. all 22 galaxies), 
we find that AGN account for $\sim$30\% of the total IR luminosity, on average, with a range of $\sim$$10-70$\% (Table \ref{tbl-4}).  
This is similar to the average AGN contribution of $\sim$30\% we find among the hard band X-ray detected sources.  
Interestingly, we also find that the average contribution to the total IR luminosity by AGN is also $\sim$30\% for both sub-samples of SMGs and non-SMGs.  
This contribution by the AGN to the IR luminosity of SMGs is a factor of $\sim$2 times larger than that reported by \citet{ap08a} which we attribute to the differences in the IR luminosities arising from differences in the SED fitting methods; as previously stated, \citet{ap08a} included the radio photometry in their SED fitting leading to IR luminosities typically larger than those presented here.  

In the right panel of Figure \ref{fig-8} we plot AGN luminosity versus the difference between the 24~$\micron$ and best-fit estimate IR luminosities.  
A clear trend of increasing AGN luminosity with increasing overestimation of the IR luminosity using 24~$\micron$ photometry alone is observed;  or, in other words, AGN luminosity appears to scale with increasing mid-infrared luminosity.  
Performing an ordinary least squares fit to the data, we find that 
\begin{equation}
\label{eq-agnfrac}
\log\left(\frac{L_{\rm IR}^{\rm AGN}}{L_{\sun}}\right) = 
	(0.50 \pm 0.04) \log \left(\frac{L_{\rm IR}^{24} - L_{\rm IR}^{\rm all}}{L_{\sun}}\right) + (5.61 \pm 0.52).
\end{equation}
While the differences between the 24~$\micron$ and best-fit IR luminosities are large for galaxies having large (i.e. $\ga$60\%) mid-infrared AGN fractions, we find these quantities to be unrelated for galaxies having smaller mid-infrared AGN fractions.  
We also find that the AGN luminosity accounts for only 16\% of the difference between the 24~$\micron$ derived and best-fit IR luminosities, on average.    
This indicates that the AGN contribution to the mid-infrared excesses, which would result in overestimates of IR-based SFRs, is close to negligible when compared to the improper bolometric correction applied when estimating total IR luminosities by fitting local SED templates with 24~$\micron$ data alone.  
As shown in $\S$ 4.4, SFRs derived by subtracting the AGN fraction from the 24~$\micron$ photometry are larger than the AGN corrected best-fit IR SFRs by a factor of $\sim$1 to $\sim$3.3, on average for redshifts below and above $z\sim1.4$, respectively.  
Furthermore, the AGN-corrected 24~$\micron$ SFRs are factors of $\sim$1 to $\sim$68 times larger than the corresponding UV-corrected SFRs for sample galaxies at redshifts above $z\sim1.4$ .  
Therefore, the AGN is not the dominant source of the inferred ``mid-infrared excesses" among these systems.  

Since the FIR-radio correlation is thought to solely arise from the processes associated with massive star formation, we looked to see if the IR/radio ratios themselves are sensitive to the fractional AGN power by comparing the ratio of $L_{\rm IR}^{\rm AGN}/L_{\rm IR}^{\rm all}$ with $q_{\rm IR}$ values.  
A trend was not found suggesting that $q_{\rm IR}$ is insensitive to the fractional IR output by the AGN.   
This result likely arises because the AGN contributes to the radio and infrared emission by different amounts.  

\subsubsection{The Persistence of  ``Mid-Infrared Excess" Sources}
By subtracting off our estimate of the AGN contribution to the total IR luminosity, we determine whether this is enough to account for the remaining IR excesses for galaxies in Figure \ref{fig-7}.  
We recreate the top and middle panels of Figure \ref{fig-7} in the top and bottom panels of Figure \ref{fig-9}, respectively, by computing the 24~$\micron$-derived and best-fit IR-based SFRs after subtracting off the estimated AGN contribution for the 8 galaxies having optical detections which allowed for proper estimates of extinction corrected UV SFRs;  
the decrease in IR SFRs are indicated by vertical lines.  

Among these sources we find that \((L_{\rm IR}^{24} - L_{\rm IR}^{\rm 24,noAGN})/L_{\rm IR}^{24}\) and \((L_{\rm IR}^{\rm all} - L_{\rm IR}^{\rm all,noAGN})/L_{\rm IR}^{\rm all}\) are both $\sim$45\%, on average.   
However, even by using the AGN-corrected 24	~$\micron$ SFRs, none of the previously identified mid-infrared excess sources are removed.    
While the AGN fractions are similar in both cases, the AGN-subtracted 24~$\micron$ SFRs are still $\sim$6 times larger, on average, than the AGN-corrected SFRs using the best-fit IR luminosities.  
In the bottom panel of Figure \ref{fig-9} we find that by subtracting the AGN luminosities we are able to account for the observed IR excesses in only one more source;
GN-IRS~12, which is plotted as an upper limit, is now no longer considered to have excess IR emission.  
Therefore, of the 8 sources with optical detections such that we can estimate a UV extinction correction in the redshift range between $1.4\la z \la 2.6$, 6 of which would be classified as ``mid-infrared excesses" using 24~$\micron$-based IR SFRs,  3 (GN-IRS 2, 15 and 17) remain mid-infrared excess sources even after subtraction of the AGN component.

Ultimately, we find that even after correcting for the presence of an AGN, the IR-based SFRs are a factor of $\sim$2.8 times larger, on average, than those derived from extinction corrected UV measurements for $1.4 \la z\la 2.6$ galaxies with secure estimates of their rest-frame UV slopes.     
Or, in other words, AGN are only able to account for $\sim$35\% of the excess IR emission measured, on average.    
Since we do not believe our IR luminosities are overestimated by such a large amount, given the small uncertainties associated with our SED fitting,  
we interpret this result to suggest that using the UV slope to derive an extinction correction is not appropriate for the entire sample; extinctions are underestimated thereby yielding underestimates for the UV-derived SFRs.    
This may not be too surprising given that there are catastrophic failures using the rest-frame UV slope to derive the extinction for galaxies having both low and high extinctions.  
If extinction within a galaxy is too high, this method will fail simply because the galaxy's ISM will become optically thick.  
In the case of low extinction the UV slope will no longer be as sensitive to the amount of extinction since variations arising from contributions by a galaxy's old stellar population will become important \citep{nr06}.      

\subsubsection{AGN Dominated Sources}
Of the remaining 3 mid-infrared excess sources after correcting for the contribution of the AGN to the total IR luminosity, 
GN-IRS~17 is the only source  plotted as an upper limit, however all have mid-infrared AGN fractions $\ga$60\% suggesting that they are AGN dominated.  
While the remaining IR excess for GN-IRS~17 may arise from improper SED fitting due to only having upper limits at 70 and 850~$\micron$, we note that its optical morphology is point-like and consistent with being an AGN.  
It may then be that the SED templates used for the fitting, which are based on star-forming galaxies, are not appropriate resulting in an incorrect estimate of the total IR luminosity.  
In fact, this source has recently been identified as a Compton-thick AGN by \citet{da08a}.  
Similarly, GN-IRS~2 is thought to be a Compton-thick AGN \citep{ap08a}, so it is not surprising to find that it lies above the threshold even though it is not plotted as an upper limit.
Thus, the only remaining mid-infrared excess source which is not plotted as an upper limit or has not been previously identified as a Compton-thick AGN is GN-IRS~15.

Inspecting the IRS spectra for this source, it shows strong evidence of a deep silicate absorption feature and weak PAH equivalent widths, suggestive of a deeply embedded AGN.  
The ACS imagery of this source is also point-like, which too seems consistent with the source being an AGN.  
Consequently, we believe that GN-IRS~15 is primarily powered by an AGN (i.e. a Compton-thick AGN) for which our SFR estimates are not physically meaningful.      

The other source which appears to be clearly AGN dominated according to its mid-infrared AGN fraction is GN-IRS~21.  
We note that this source never appears to be a mid-infrared excess source even when comparing its UV corrected SFR to that derived from SED fitting 24~$\micron$ photometry.   
However, by attempting to correct for the AGN in this source, we find that the best-fit IR derived SFR becomes $\sim$5 times smaller than that of the UV corrected SFR; 
clearly the UV and IR SFR estimates for this AGN dominated source are not physically meaningful.  
We also note that GN-IRS~4 and 12, which just barely drop below the mid-infrared excess threshold in the bottom panel of Figure \ref{fig-9}, have mid-infrared AGN fractions near 50\%.  
However, since our mid-infrared AGN fractions are considered to be upper limits, we believe these systems to likely be dominated by star formation.   

Of the 6 originally identified "mid-infrared excess" sources, based on extrapolating IR-based SFRs from SED fitting 24~$\micron$ photometry alone, 3 remain even after accounting for the presence of AGN and applying proper bolometric corrections.  
This suggests that the sky and space densities of Compton-thick AGN reported by \citet{ed07b} likely overestimate the true values by a factor of $\sim$2, yielding corrected values of $\sim$1600~deg$^{-2}$ and $\sim$$1.3 \times10^{-4}$~Mpc$^{-3}$, respectively.


\subsubsection{Are ``Mid-Infrared Excess" Sources DOGs?}
\begin{figure*}
\plotone{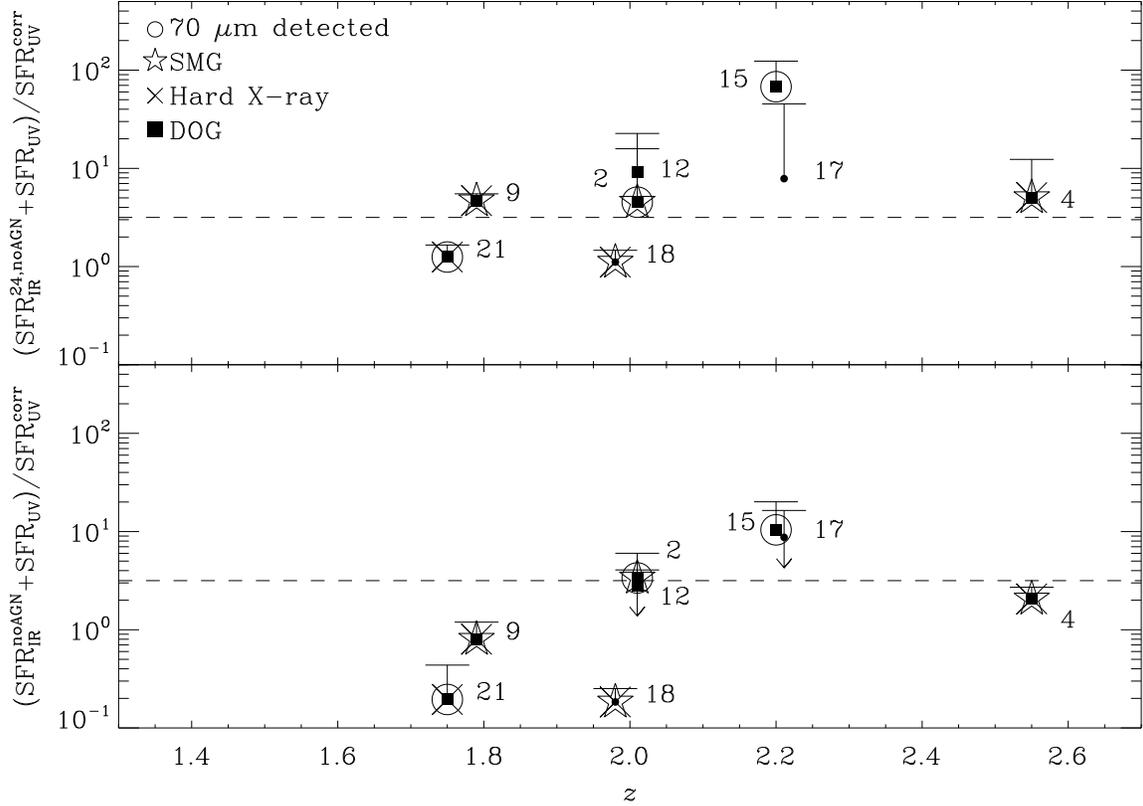}
\caption{
The top panel is similar to that of the top panel of Figure \ref{fig-7}, except that we have subtracted the AGN fraction from the 24~$\micron$ photometry before calculating the associated IR luminosities and SFRs.  
We also only show galaxies for which a proper extinction corrected UV SFR could be calculated.    
The decrease, which is $\sim$45\% on average among these sources, is shown by a {\it vertical} line.  
In the bottom panel we subtract the AGN contribution to our best estimates of the IR luminosity  (i.e. $L_{\rm IR}$ values estimated from SED fitting all available data to derive SFRs).  
The decrease in SFR by correcting for the AGN ($\sim$45\% on average among these sources as well) is again shown by a {\it vertical} line.  
While the decrease to the SFRs by correcting for AGN is similar among these sources, the AGN-subtracted 24~$\micron$ SFRs are still $\sim$6 times larger, on average, than the AGN-corrected SFRs from our best-fit IR luminosities.  
\label{fig-9}}
\end{figure*}

In Figure \ref{fig-9} we label sources which would be identified as dust-obscured galaxies (DOGs) using the definition of \citet{ad08}, $f_{\nu} (24~\micron)/f_{\nu} (R) > 982$, to see if they clearly separate out as mid-infrared excess sources.  
We estimate $R$ band flux densities by interpolating between the $V$ and $i_{775}$ band photometry.  
More than half of the sample (i.e. 13 sources, 7 of which we could not measure proper extinction corrected UV SFRs and are therefore not included in Figure \ref{fig-9}) fit the DOG definition.  
Not all DOGS can be classified as mid-infrared excess sources.  
Conversely, not all mid-infrared excess sources are DOGs. 
This result is not too surprising since DOGs are thought to be powered by both star formation and AGN processes \citep{ap08b}.  
However, we do find that DOGs are only identified for sources wtih $z > 1.4$.  
In fact, 87\% of our sample having a redshift greater than 1.4 are classified as DOGs.  

\begin{figure}
\plotone{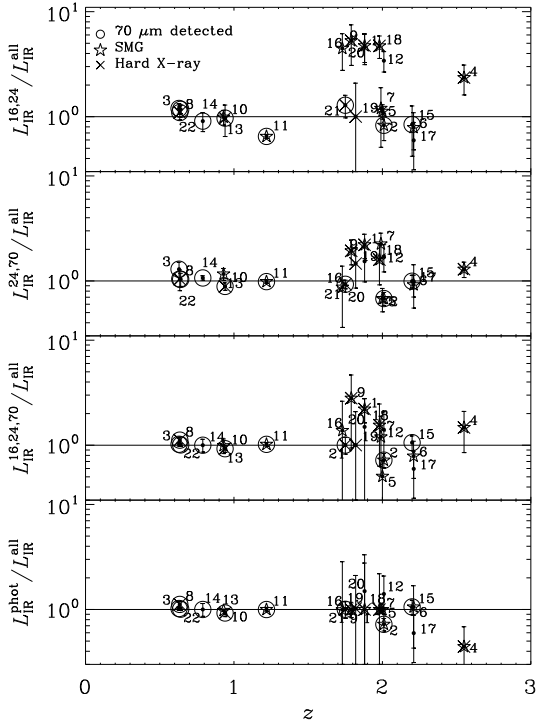}
\caption{
In the top panel we plot the ratio between the 16 and 24~$\micron$-derived IR luminosity ($L_{\rm IR}^{16,24,70}$) to that derived using the additional constraints from the IRS spectroscopy and 850~$\micron$ submillimeter data.  
Similarly, in the second and third panels, we plot the ratio of the 24 and 70~$\micron$-derived ($L_{\rm IR}^{24,70}$) and the 16, 24, and 70~$\micron$-derived  ($L_{\rm IR}^{16,24,70}$) IR luminosities to those estimated from fitting all available data, respectively.     
In the bottom panel we plot the ratio between IR luminosities derived from fitting all available photometric data ($L_{\rm IR}^{\rm phot}$ to our best-fit IR luminsoties, which also included the IRS spectra in the SED template fitting.  
This comparison is shown to see how well the mid-infrared spectral region is characterized by fitting photometry alone and how differences are reflected in the associated IR luminosity values.  
Errors in the fitting are shown.    
In each panel a {\it solid} line is plotted indicating a ratio of unity.   
\label{fig-10}}
\end{figure}

\begin{figure}
\plotone{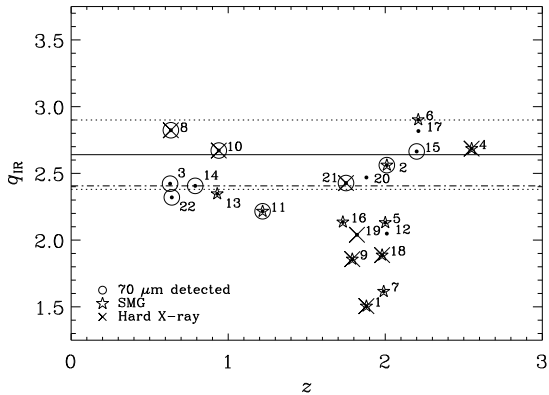}
\caption{
We plot $q_{\rm IR}$ ratios, as defined in Equation \ref{eq-qIR}, calculated using IR luminosities from SED fitting all available data against redshift.  
The {\it solid} line indicates the local average $q_{\rm IR}$ value of 2.64 \citep{efb03} and the {\it dotted} lines are the $\pm$1$\sigma$ values.
The thick {\it dot-dash} line indicates the median value for the sample and falls just above the $-1$-$\sigma$ line. 
\label{fig-11}}
\end{figure}

\begin{figure}
\plotone{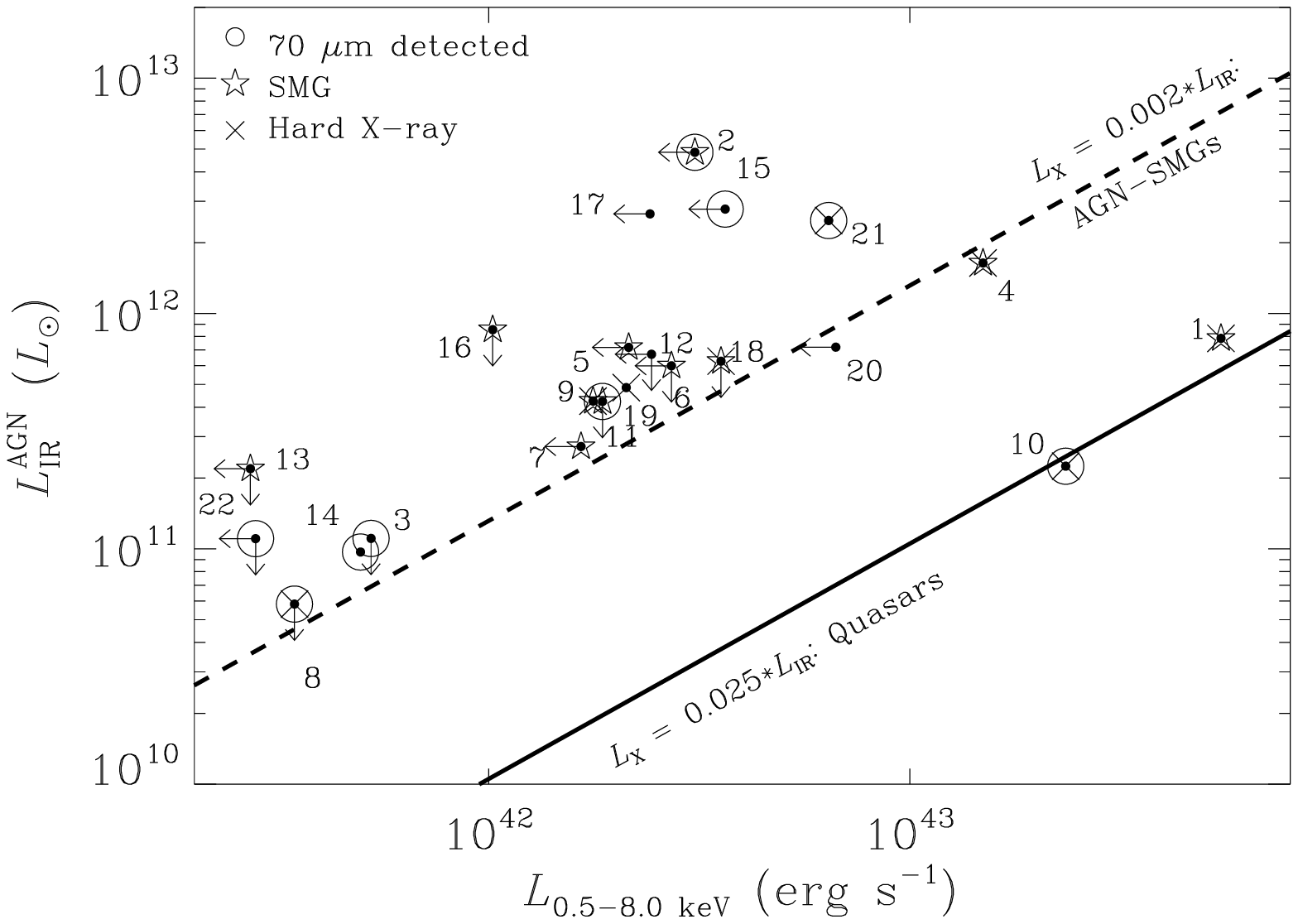}
\caption{
The estimated AGN contribution to the total IR luminosites are plotted against the $0.5-8.0$~keV luminosities.  
Upper limits limits to the AGN and X-ray luminosities are identified by appropriately directioned arrows.  
The {\it solid} line indicates the canonical relation between the IR and X-ray luminosities of AGN-dominated sources (i.e. quasars) reported by \citet{me94}.  
The {\it dashed} line indicates the fit to AGN-classified SMGs presented in \citet{da05} and assumes an AGN-IR luminosity fraction of 100\%.   
The 3 Compton-thick AGN (i.e. GN-IRS~2, 15, and 17) have dust covering fractions significantly larger than that of typical quasars.     
\label{fig-12}}
\end{figure}

\subsection{Application For Other Fields with Deep {\it Spitzer} and {\it Herschel} Far-Infrared Imaging}
We compare the results of our SED fitting to situations when only mid-infrared photometry (e.g. at 16 and/or 24~$\micron$) along with additional measurements at 70~$\micron$ are available.      
Such a comparison is especially important for the FIDEL fields, because of their deep 70~$\micron$ imaging, as well as for upcoming dedicated {\it Herschel} programs such as GOODS-{\it Herschel} (GOODS-H; PI: D.~Elbaz).  

In the third panel of Figure \ref{fig-10} we plot the ratio of IR luminosities derived by fitting the 16, 24, and 70~$\micron$ photometry to $L_{\rm IR}^{\rm all}$ as a function of redshift.    
The median ratio has a value of $\sim$$1.02\pm 0.51$.
We find the ratio to be approximately constant with increasing redshift, with average values of $\sim$$1.01\pm0.06$ and $\sim$$1.20\pm0.61$ for redshifts below and above $z \sim 1.4$, respectively, suggesting that the systematic variations introduced by the presence of strong PAH emission redshifing into the 24~$\micron$ band at $z\ga1.4$ is largely taken into account by having the additional 70~$\micron$ data point.  
We also see that the dispersion significantly increases with redshift.  
We therefore believe that using mid-infrared photometry with the addition of a 70~$\micron$ measurement provides fairly reliable estimates for the true IR luminosity 
compared to when fitting with the 24~$\micron$ data alone.     
We also note that it is the addition of the 70~$\micron$ data, and not the additional 16~$\micron$ photometry,  which is responsible for the improved determination of IR luminosities.  

In the first (top) panel of Figure \ref{fig-10} we plot the ratio of IR luminosities derived by fitting the 16 and 24~$\micron$ photometry to those using all of the available data versus redshift.    
The median ratio of $16+24~\micron$-derived to our best-fit IR luminosities is $\sim$1.17$\pm 1.65$.   
As with the comparison made above, the median and dispersion in the ratio of IR luminosity estimates increases with redshift, going from $\sim$$0.99\pm0.19$ to  
$\sim$$1.29\pm1.82$ at redshifts below and larger than $z\sim1.4$.  
This is of course not surprising since fitting the 16 and 24~$\micron$ photometry for redshifts beyond $z\sim1$ is complicated by the presence of PAH and silicate absorption features being redshifted into both bands.  
Instead, taking the ratio of $24+70~\micron$-derived to our best-fit IR luminosities, we find a median ratio of $\sim$1.07$\pm 0.45$ which is illustrated in the second panel of Figure \ref{fig-10}.    
Similarly, we find the median ratio to remain near unity  for galaxies below and above $z\sim 1.4$, however the increase in the dispersion between these subsets is less significant being $\sim$$1.04\pm0.13$ and $\sim$$1.29\pm0.53$, respectively.  
Furthermore, we find that the additon of the 16~$\micron$ photometry to SED fitting the 24 and 70~$\micron$ data does not result in largely different IR luminosity estimates; 
the median ratio of IR luminosities from fitting the $16-70~\micron$ data to those from fitting only the 24 and 70~$\micron$ data is $\sim$$1.03\pm 0.29$ and ranges from $\sim$$1.03\pm0.12$ to $\sim$$1.03\pm0.34$ for galaxies below and above $z\sim1.4$, respectively.  
Such a comparison is useful since most deep fields only have coverage at 24 and 70~$\micron$.  
Consequently, the addition of the 16~$\micron$ photometry doesn't appear to improve the fitting significantly compared with the addition of the 70~$\micron$ data.  

It is interesting to see how including the mid-infrared spectra affects the SED template fitting compared to when using only the available photometry.  
We plot the ratio of $L_{\rm IR}^{\rm phot}/L_{\rm IR}^{\rm all}$ as a function of redshift in the bottom panel of Figure \ref{fig-10}.  
We find that the ratio has a median that is approximately unity and a dispersion of $\pm 0.22$.    
While the median of the ratio remains near unity for redshifts below and above $z\sim1.4$, we again find that the dispersion increases with redshift, taking values of $\sim$0.06 and $\sim$0.26 in these redshifts bins, respectively.  

Furthermore, we find that using the 70~$\micron$ data alone to derive IR luminosities does not appear to be reliable as this method significantly underestimate the true IR luminosity.  
Among the 70~$\micron$ detected sources (only 2/9 which are SMGs), the best-fit  IR luminosities are larger by an average factor of $\sim$$1.9\pm 0.22$ than the IR luminosities derived from the 70~$\micron$ data alone.  
Such a result is expected if the average galaxy at high redshift has an SED which is cooler than that of a a local galaxy at the same luminosity \citep[e.g.][]{ap06,ak06,mh07}.  
However, to make a conclusive statement about such an effect requires a precise sampling of the full IR SED, including the peak, which we do not have.  

Seeing that the addition of 70~$\micron$ data, alone, to mid-infrared photometry yields relatively reliable estimates of IR luminosity among these galaxies, FIDEL observations (i.e. deep 70~$\micron$ imaging) of the entire GOODS-N field are used to better constrain estimates of the IR luminosity for all 24~$\micron$ detected sources having spectroscopic redshifts in E.J. Murphy et al. (2009, in preparation).  
By doing so, we are able to improve estimates for how the IR luminosity density evolves with redshift and characterize better the number density of mid-infrared excess sources.  

\begin{deluxetable*}{ccl}
\tablecaption{Summary of Sources \label{tbl-6}}
\tablewidth{0pt}
\tablehead{
\colhead{ID} & \colhead{Class$^{a}$} & \colhead{Comment}
}
\startdata
 1&	AGN       &   UV-slope indeterminate\\
 2&   CT-AGN &  Mid-infrared excess even after AGN subtraction and proper bolometric correction\\
 3& 	SF           &   Strong PAH Features; UV-slope indeterminate\\
 4&	SF           &   Strong PAH features; UV-slope underestimates extinction\\
 5&	SF           &   UV-slope indeterminate\\
 6&	SF           &   UV-slope indeterminate\\
 7&	SF           &   Strong PAH features; UV-slope indeterminate\\
 8&	SF           &   Strong PAH features; UV-slope indeterminate\\
 9&	SF           &  UV-slope correctly estimates extinction\\
10&	SF           &  UV-slope indeterminate\\
11&	SF           &  Strong PAH features; UV-slope correctly estimates extinction\\
12&	SF           &  Strong PAH features; UV-slope underestimates extinction; low $q_{\rm IR}$ value\\
13&	SF           &  UV-slope overestimates extinction \\
14&	SF           &  UV-slope correctly estimates extinction\\
15&  CT-AGN & Mid-infrared excess even after AGN subtraction and proper bolometric correction\\
16&	SF           & Strong PAH features \\
17&  CT-AGN & Mid-infrared excess even after AGN subtraction and proper bolometric correction\\
18&	SF           & Strong PAH features; UV-slope overestimates extinction\\ 
19&	SF           &  Strong PAH features\\
20&	SF           &   UV-slope indeterminate\\
21&  AGN       &  UV-slope \& 24~$\micron$ conspire to produce consistent SFRs \\
22&	SF           &  Strong PAH features; UV-slope indeterminate
\enddata
\tablecomments{$^{a}$ Final classification: AGN - IR AGN fraction larger than 50\%; CT-AGN - Compton-thick AGN; SF- star formation dominated source}
\end{deluxetable*}

\subsection{The FIR-Radio Correlation}
Using these data also enables us to look at the behavior of the FIR-radio correlation among a diverse group of galaxies spanning a redshift range between $0.6 \la z \la 2.6$.  
In calculating the radio based IR luminosities and associated SFRs we have assumed that the FIR-radio correlation remains constant with redshift.  
We will now try to verify if such an assumption was appropriate.  

\subsubsection{Lack of Evolution with Redshift}
In Figure \ref{fig-11} we plot $q_{\rm IR}$, defined in Equation \ref{eq-qIR}, as a function of redshift.
Infrared luminosities are estimated from SED fitting using all available photometry and the IRS spectroscopy.  
The median $q_{\rm IR}$ value, plotted as {\it dot-dashed} line in Figure \ref{fig-11}, is $2.41 \pm0.39$~dex, below, but consistent with, the local value of $2.64\pm0.26$  \citep{efb03}.  
Interestingly, we note that the 4 galaxies having the largest mid-infrared AGN fractions (GN-IRS~2, 15, 17, and 21) have $q_{\rm IR}$ values very near the canonical ratio, being within 0.18~dex, on average.   
This indicates that the FIR-radio correlation is a poor discriminant of obscured AGN activity.   

While the correlation appears to remain linear with redshift for the few objects being studied here, 
more than 50\% of the SMGs in this sample (i.e. 7/11) have IR/radio ratios $<2.24$.  
This is a factor of $\ga$2.5 below the average value measured in the local Universe for both normal star-forming galaxies and ULIRGs.  
These galaxies drive the large dispersion found among the entire sample relative to what is found in the local Universe.    
The SMGs show a scatter in $q_{\rm IR}$ of 0.44~dex compared to 0.27~dex for the non-SMGs suggesting a more heterogeneous population.  
In contrast, 12/13 ULIRGs in the \citet{yun01} sample, (i.e. excluding a single, radio loud AGN), have a median IR/radio ratio which is 33\% larger than the canonical value and a dispersion of only 0.28~dex.        
While the average $q_{\rm IR}$ value for the non-SMGs is 2.43~dex, nearer to the local value of 2.64~dex, the median $q_{\rm IR}$ value among all SMGs in the sample is 2.14~dex, a factor of $\sim$3.2 times lower than the canonical value.  

The errors in $L_{\rm IR}^{\rm all}$ appear too small to explain this systematic departure from the canonical IR/radio ratio among the SMGs;  
by increasing $L_{\rm IR}^{\rm all}$ values by the 3$\sigma$ errors and recomputing $q_{\rm IR}$, the median value among the SMGs is 2.16~dex, still a factor for $\sim$3. times lower than the canonical ratio.
This result is consistent with the findings of \citet{ak06} who reported similarly low IR/radio ratios for a sample of $\approx$15 SMGs out to $z\sim3.5$ detected at 350 and 850~$\micron$.  
The possible reasons for the observed systematic departure off the FIR-radio correlation for this galaxy population are discussed below. 

\subsection{Physical Explanations for Differences in $q_{\rm IR}$ among the SMGs}
Assuming that the uncertainty in our estimates of IR luminosity are not responsible for the low IR/radio ratios among the SMGs, there are a number of physical explanations for why galaxies may have IR/radio ratios that are significantly lower than the local correlation value.  
First we note that a single spectral index was used to $K$-correct all of the observed radio flux densities. 
While we have assumed a radio spectral index of $\alpha \approx 0.8$, it would take a negative slope (i.e. $\alpha \approx -0.4$) to return the average IR/radio ratio back to the canonical value for the SMGs. 
Flat or inverted spectra associated with such indices are indicative of AGN (e.g. gigahertz peaked sources arising from synchrotron self absorption). 
The radio spectral slopes for $z\sim2$ ULIRGs are typically not found to exhibit such indices; 
\citet{as08} report only 3/48 {\it Spitzer} selected ULIRGs have radio spectral indices between 610~MHz and 1.4~GHz (i.e. $\alpha^{\rm 610~MHz}_{\rm 1.4~GHz}) <  0.4$.  
Each of these sources also show very weak PAH emission which is unlike the SMGs presented here.  

\citet{ak06} attributed the low IR/radio ratios to a possible decrease in the effective dust emissivity index for SMGs compared to local star-forming galaxies.  
However, this is hard to prove without a fine sampling of the full IR SED.  
The simplest explanation is extra radio emission due to the presence of an AGN.  
Optical spectroscopy of the SMGs indicate the presence of AGN activity \citep[e.g.][]{as04,sc05}.  
Analysis of the X-ray data on SMGs also 
indicates the presence of an AGN in most SMGs \citep{da05}. 
Thus, it is very likely that some of the radio emission arises from the nuclear black hole as can be seen by the fact that a number of the most deviant $q_{\rm IR}$ values are in fact detected in the hard band ($2.0-8.0$~keV) X-rays.  

In order to assess if the departure from the canonical $q_{\rm IR}$ value for the SMGs is due to radio emission from an AGN, 
we recompute $q_{\rm IR}$ ratios by first subtracting the AGN component of the IR luminosity and attempting to make a similar correction to the radio.  
This is done by assuming that, like the IR emission, the radio emission of Mrk~231 is dominated by its AGN.   
This is reasonable since Mrk 231 has a $q_{\rm IR} =  2.39$, nearly a factor of $\sim$2 below the nominal value for star-forming galaxies.
Taking the 20~cm flux density of Mrk~231 of 0.309~Jy \citep{ccb02} along with its redshift ($z=0.042$), the corresponding 20~cm specific luminosity is $3.36\times10^{-3} L_{\sun}~{\rm Hz}^{-1}$.  
The fraction of radio power associated with the AGN for each galaxy was then estimated by scaling the radio luminosity of Mrk~231 by the mid-infared AGN fractions of the mid-infrared luminosity  obtained in $\S$\ref{sec-AGNfrac}.  
This contribution is then subtracted from the measured 20~cm specific luminosities for each galaxy before computing the new IR/radio ratios.  

Rather than decreasing the dispersion among the sample galaxies, this attempt to remove emission arising from the AGN has actually increased the dispersion from 0.49 to  0.54 dex.  
This result suggests that the fractional infrared output of the AGN does not scale with the fractional output in the radio, consistent with what was found in $\S$5.1 by comparing the fractional AGN power with $q_{\rm IR}$.  
{

Another explanation for the low IR/radio ratios may be related to galaxy-galaxy interactions.    
The morphology of SMGs seem to suggest major mergers which are driving intense bursts of star formation \citep[e.g.][]{sc03}.   
In the local Universe, a number of interacting galaxy pairs (i.e. so called ``Taffy" galaxies) exhibit IR/radio ratios which are a factor of $\sim$2 lower than the canonical value  \citep{chss93,chj02}.  
These systems are characterized by a synchrotron "bridge", a region of bright radio continuum connecting the galaxy pairs which is thought to arise from a recent interaction and likely contain both cosmic rays and magnetic fields.    
The excess radio emission arising from the gaps between these galaxy pairs typically accounts for half of the total radio emission from the entire system.  
If SMGs have indeed undergone a similar major-merger event, leading to molecular gas and synchrotron bridges between galaxy pairs which do not yet facilitate a significant amount of ongoing star formation, this scenario may explains the lower IR/radio ratios.

On the other hand, perhaps the extreme star-bursts occurring within SMGs lead to physical conditions which affect the acceleration and cooling processes of cosmic-ray electrons which boost the synchrotron emissivity.    
The sample of SMGs presented here is far too small to properly address this question, which will be discussed in a future paper.

\subsection{IR and X-ray Luminosities}
A comparison of the AGN and $0.5-8.0$~keV luminosities are shown in Figure \ref{fig-12} for the entire sample.  
There exists a loose relationship between the IR and full band X-ray luminosities of AGN dominated sources (i.e. quasars) such that $L_{\rm 0.5-8.0~keV}/L_{\rm IR} \sim 0.025$ \citep[e.g.][]{me94} as indicated by the {\it solid} line in Figure \ref{fig-12}.  
We also include the fit to AGN-classified SMGs presented in \citet{da05} having an average $L_{\rm 0.5-8.0~keV}/L_{\rm IR}$ ratio of $\sim$0.002.  

When the AGN luminosities are plotted against the full band X-ray luminosities we find that nearly every galaxy, save GN-IRS~1 and GN-IRS~10, remain either near or above the AGN-classified SMG relation and well away from the canonical AGN relation line for quasars.  
Despite the fact that GN-IRS~1 and GN-IRS~10 are hard band X-ray detected, they have  mid-infrared AGN fractions which are $\la$50\%.  
In fact, the median mid-infrared AGN fraction among all hard band X-ray detected sources is $\sim$44\% with a range of $\sim$$23-62$\%.  
This indicates that a hard band X-ray detection does not necessarily imply a galaxy is AGN dominated.       

We believe that the reason most sources lie generally above the above the AGN-classified SMG relation is primarily from the overestimation of AGN luminosities; 
as previously stated, the mid-infrared AGN fractions and associated AGN luminosities should be thought of as upper limits since hot dust contributing to the mid-infrared continuum emission may be powered by star formation.    
Alternatively, there is some uncertainty on the absorption corrections to the X-ray flux, and it is possible that they could be a factor of $\sim$2 higher \citep{da08b}.  
It is also possible that the AGN in these sources contribute significantly more to the total IR output than the few percent suggested for SMGs by \citet{da05} arising from a large AGN dust covering factor.  
The 4 galaxies having the largest mid-infrared AGN fractions (GN-IRS~2, 15, 17, and 21), three of which are thought to be Compton-thick AGN (GN-IRS~2, 15, and 17), all lie well away from the typical quasar relation suggesting that they have significantly larger dust covering fractions than quasars.     

\section{Conclusions}
In the present study we have used observations from the mid-infrared to the submillimeter to properly characterize the IR luminosities for a diverse sample of 22 galaxies spanning a redshift range of $0.6 \la z \la 2.6$.  
In addition, we have used the mid-infrared spectra of these sources to estimate the fractions of their IR luminosities which arise from an AGN.  
Our conclusions can be summarized as follows:


\begin{enumerate}

\item
IR ($8-1000~\micron$) luminosities derived by SED fitting observed 24~$\micron$ flux densities alone are well matched to those when additional mid-infrared spectroscopy and 16, 70, and 850~$\micron$ photometry are included in the fits for galaxies having $z \la 1.4$ and $L_{\rm IR}^{\rm 24}$ values typically  $\la$$3\times10^{12}~L_{\sun}$.  
In contrast, for galaxies lying in a redshift range between 1.4 and 2.6 with $L_{\rm IR}^{\rm 24}$ values typically $\ga$$3\times10^{12}~L_{\sun}$, IR luminosities derived by SED template fitting using observed 24~$\micron$ flux densities alone overestimate the true IR luminosity by a factor of $\sim$5, on average, compared to fitting all available data.  
A comparison between the observed mid-infrared spectra with that of the SEDs chosen from fitting 24~$\micron$ photometry alone and from fitting all available photometric data demonstrates that local high luminosity SED templates show weaker PAH emission by an average factor of $\sim$5 in this redshift range and do not properly characterize the contribution from PAH emission.   

\item
After decomposing the IR luminosity into star forming and AGN components, we find the AGN luminosity to be increasing with increasing difference between the 24~$\micron$-derived and our best-fit IR luminosities.  
Such a trend suggests that the AGN power increases with mid-infrared luminosity.    
However, we also find that the median fraction of the AGN to the difference between the 24~$\micron$-derived and best-fit IR luminosities is only 16\% suggesting the AGN power is almost negligible compared to the bolometric correction necessary to properly calibrate the 24~$\micron$-derived IR luminosities.  

\item 
Infrared luminosities calculated using both mid-infrared and 70~$\micron$ photometry agree to within $\sim$50\%, on average, to those calculated using additional mid-infrared spectroscopic and submillimeter data and do not show any systematic deviations with increasing redshift.  
Including the submillimeter photometry in this calculation improves the agreement between the two sets of IR luminosities to within $\sim$20\%.  
This result implies that IR luminosities derived for deep fields having 70~$\micron$ data, like the FIDEL fields, should be able to obtain fairly reliable estimates for the total IR luminosity of sources up to $z\sim3$ even in the absence of submillimeter data.  
This will be investigated for GOODS-N in a forthcoming paper.

\item 
Even after correcting for the presence of an AGN, IR-based SFRs are still larger than those derived from extinction corrected UV measurements by a factor of $\sim$2.8, on average, for the $1.4 \la z\la 2.6$ galaxies with secure estimates of their rest-frame UV slopes.
Or, in other words, the AGN emission is only able to account for $\sim$35\% of the``excess" IR emission, on average, and is not the dominant cause of the mid-infrared excesses observed in these systems.     
This suggests that either the SED fitting is overestimating the total IR luminosity, which seems unlikely given the small uncertainties associated with our fitting, 
or perhaps that the UV extinction correction underestimates the true extinction factors for these galaxies.  

\item
By using proper bolometric corrections and correcting for the presence of AGN, we are able to account for half of the sources which are identified to have a "mid-infrared excess", as defined by \citet{ed07a},  based on their 24~$\micron$-derived IR luminosities.  
This indicates that the sky and space densities of Compton-thick AGN reported by \citet{ed07b} are likely high by a factor of $\sim$2.  
We therefore report corrected sky density for Compton-thick AGN of $\sim$1600~deg$^{-2}$, and a corresponding space density of $\sim$$1.3 \times10^{-4}$~Mpc$^{-3}$.    

\item
We do not see any clear signatures of evolution in the FIR-radio correlation with redshift out to $z\sim2.6$ for this sample of  galaxies.  
However, the median IR/radio ratio measured for the SMGs included in our IRS selected sub-sample is 2.14~dex, a factor of $\sim$3 lower than what is found locally for star-forming galaxies.  

\item
A comparison of the estimated AGN and full band ($0.5-8.0$~keV) X-ray luminosities indicate a higher IR output than observed canonically for quasars for the majority of X-ray detected sources in the sample.  
This can be explained if the AGN luminosities are being overestimated due to star formation contributing significantly to the hot dust continuum in the mid-infrared or if the AGN dust covering factor is large in these sources.  
\end{enumerate}

\acknowledgments
We would like to than M.T.~Huynh and D.~Frayer for useful discussions.  
We would also like to thank the anonymous referee for useful suggestions which helped to improve the paper.  
D.~M.~A. acknowledges funding from the Royal Society.    
This work is based on observations made with the Spitzer Space Telescope, which is operated by the Jet Propulsion Laboratory, California Institute of Technology, under a contract with NASA. Support for this work was provided by NASA through an award issued by JPL/Caltech.

\clearpage
\begin{landscape}
\setcounter{table}{2}
\begin{deluxetable*}{ccccccccccccc}
\tablecaption{Multiwavelength Photometry \label{tbl-3}}
\tablewidth{0pt}
\tablehead{
\colhead{} & 
\colhead{$f_{\rm 0.5-8.0~keV}/10^{-17}$} & \colhead{$f_{\rm 2.0-8.0~keV}/10^{-17}$} & \colhead{} & 
\colhead{$f_{\nu}(B)$} & \colhead{$f_{\nu}(V)$} & \colhead{$f_{\nu}(i)$} & \colhead{$f_{\nu}(z)$} & 
\colhead{$f_{\nu}(16~\micron)$} & \colhead{$f_{\nu}(24~\micron)$} & \colhead{$f_{\nu}(70~\micron)$} & \colhead{$f_{\nu}(850~\micron)$} & 
\colhead{$S_{\nu}(20~{\rm cm})$}\\ 
\colhead{ID} & 
\colhead{(${\rm erg~cm^{-2}~s^{-1}}$)} & \colhead{(${\rm erg~cm^{-2}~s^{-1}}$)} &\colhead{$\Gamma^{a}$} & 
\colhead{($\mu$Jy)} & \colhead{($\mu$Jy)} & \colhead{($\mu$Jy)} &  \colhead{($\mu$Jy)} &
\colhead{($\mu$Jy)} & \colhead{($\mu$Jy)} & \colhead{($\mu$Jy)} &  \colhead{($\mu$Jy)} &
\colhead{($\mu$Jy)} 
}
\startdata
  1&    1120.0&      1050.0&   0.44            &     $<$ 0.027&   $<$ 0.048&     $<$ 0.048&     $<$  0.13&           \nodata&     374&     $<$  3000&          5400&        201.89\\
  2& $<$  19.9&   $<$  34.1&   1.40$^{\dagger}$&         0.032&       0.212&         0.353&          0.71&         496.8&    1220&          4030&          7900&        127.46\\
  3&      41.0&   $<$  32.1&   1.45  $^{\ddag}$&         2.572&       3.993&        11.941&         35.84&         774.7&    1210&         11100&     $<$ 24200&        153.07\\
  4&     102.0&        79.7&   0.97            &     $<$ 0.027&   $<$ 0.048&         0.072&          0.19&          60.2&     303&     $<$  3000&          4900&         34.33\\
  5& $<$  14.0&   $<$  21.4&   1.40$^{\dagger}$&     $<$ 0.027&   $<$ 0.048&     $<$ 0.048&     $<$  0.13&          42.1&     330&     $<$  3000&          7500&        171.46\\
  6& $<$  14.4&   $<$  23.9&   1.40$^{\dagger}$&     $<$ 0.027&   $<$ 0.048&     $<$ 0.048&     $<$  0.13&     $<$  30.0&     215&     $<$  3000&          5200&         21.72\\
  7& $<$  10.9&   $<$  17.4&   1.40$^{\dagger}$&     $<$ 0.027&   $<$ 0.048&     $<$ 0.048&     $<$  0.13&     $<$  30.0&     347&     $<$  3000&          8900&        169.35\\
  8&      32.1&        25.4&   1.06            &         3.661&       5.328&        12.198&         30.46&         399.7&     721&         11100&     $<$  3500&         38.94\\
  9&     102.0&       114.0&  -0.50  $^{\ddag}$&         0.167&       0.315&         0.355&          0.72&         113.3&     414&     $<$  3000&          7700&         95.91\\
 10&     515.0&       255.0&   2.01            &     $<$ 0.027&       3.258&         5.515&         13.29&         555.8&     773&          4460&     $<$  6500&         42.69\\
 11&      28.9&   $<$  17.4&   1.63  $^{\ddag}$&         0.752&       1.073&         1.442&          6.33&         993.6&     446&         13200&          2200&        202.10\\
 12& $<$  15.7&   $<$  26.6&   1.40$^{\dagger}$&         0.091&       0.125&         0.189&          0.36&          58.8&     367&     $<$  3000&     $<$  6100&         81.82\\
 13& $<$   9.2&   $<$  14.4&   1.40$^{\dagger}$&         0.866&       1.274&         2.707&         11.35&         302.7&     367&     $<$  3000&          2100&         37.88\\
 14&      24.7&   $<$  27.3&   1.35  $^{\ddag}$&         0.188&       0.306&         2.088&         13.57&         593.9&     832&         14200&     $<$ 10600&        125.16\\
 15& $<$  19.5&   $<$  35.3&   1.40$^{\dagger}$&     $<$ 0.027&       0.061&         0.081&          0.22&         283.6&     714&          2210&     $<$  9500&         42.68\\
 16&       9.0&   $<$  12.2&   1.40            &     $<$ 0.027&   $<$ 0.048&         0.063&          0.27&         217.0&     710&     $<$  3000&          3900&        107.85\\
 17& $<$  12.8&   $<$  22.6&   1.40$^{\dagger}$&         0.700&       0.607&         0.554&          0.84&         222.7&     376&     $<$  3000&     $<$ 12000&         28.33\\
 18&      55.5&        52.1&   0.62            &     $<$ 0.027&       0.321&         0.654&          1.70&          99.6&     537&     $<$  3000&          5200&        127.64\\
 19&      42.3&        40.7&   0.51            &     $<$ 0.027&   $<$ 0.048&     $<$ 0.048&     $<$  0.13&         174.6&     512&     $<$  3000&     $<$ 13900&         83.43\\
 20& $<$  49.4&   $<$  87.8&   1.40$^{\dagger}$&     $<$ 0.027&   $<$ 0.048&         0.127&          0.36&         115.5&     498&     $<$  3000&     \nodata&         33.09\\
 21&     140.0&       132.0&   0.48            &         0.032&   $<$ 0.048&         0.171&          0.57&        1050.7&     932&          3440&     $<$  6200&         98.16\\
 22& $<$  21.5&   $<$  34.3&   1.40$^{\dagger}$&         0.540&       1.019&         5.008&         24.32&         580.6&     750&          5530&     $<$ 18200&        121.83
   \enddata
    \tablecomments{$^{a}$Photon index: $^{\dagger}$ indicates the assumed index used when only upper limits were measured;  $^{\ddag}$ indicates that the given photon index is an upper or lower limit.}
\end{deluxetable*}

\clearpage
\end{landscape}

\end{document}